\documentclass[graybox, secnum]{svmult}

\usepackage{mathptmx}       
\usepackage{helvet}         
\usepackage{courier}        
\usepackage{type1cm}        
\usepackage{makeidx}         
\usepackage{graphicx}        
\usepackage{multicol}        
\usepackage[bottom]{footmisc}
\usepackage{hyperref}        
\usepackage{soul}            
\hypersetup{colorlinks=true,urlcolor=blue}
\usepackage[square,numbers,compress]{natbib}
\makeindex             

\usepackage{xcolor}
\usepackage{amsmath,amsfonts,amssymb}
\usepackage{bbold}
\usepackage{tensor,slashed}
\usepackage{slashbox}

\newcommand{\be}{\begin{equation}}
\newcommand{\ee}{\end{equation}}

\renewcommand{\imath}{\ensuremath{\mathrm{i}}}
\newcommand{\unit}{\mathbb{1}}

\newcommand{\p}{\partial}

\newcommand{\gb}{\bar{g}}

\newcommand{\Cb}{\bar{C}}
\newcommand{\Db}{\bar{D}}
\newcommand{\Rb}{\bar{R}}

\newcommand{\cD}{\mathcal{D}}
\newcommand{\cL}{\mathcal{L}}
\newcommand{\cM}{\mathcal{M}}
\newcommand{\cO}{\mathcal{O}}
\newcommand{\cR}{\mathcal{R}}
\newcommand{\cT}{\mathcal{T}}

\newcommand{\hh}{\hat{h}}

\begin{document}
	\title*{The Functional Renormalization Group \\[1.0ex] in Quantum Gravity\thanks{ Invited chapter for the "Handbook of Quantum Gravity" (Eds.\ C.\ Bambi, L.\ Modesto and I.L.\ Shapiro, Springer Singapore, expected in 2023)}}
	\author{Frank Saueressig}
	\institute{Frank Saueressig \at Institute for Mathematics, Astrophysics and Particle Physics, Radboud University \mbox{Nijmegen}, Heyendaalseweg 135, 6525 AJ Nijmegen, The Netherlands \\ \email{f.saueressig@science.ru.nl}}
	%
	%
	\maketitle
	\abstract{The gravitational asymptotic safety program envisions a high-energy completion of the gravitational interactions by an interacting renormalization group fixed point, the Reuter fixed point. The primary tool for investigating this scenario are functional renormalization group equations, foremost the Wetterich equation. This equation implements the idea of the Wilsonian renormalization group by integrating out quantum fluctuations shell-by-shell in momentum space and gives access to the theory's renormalization group flow beyond the realm of perturbation theory. This chapter gives a pedagogical introduction to the gravitational asymptotic safety program with a specific focus on clarifying conceptual points which led to confusion in the past. We provide a step-by-step introduction to the Wetterich equation and its most commonly used non-perturbative approximations. This exposition also introduces recent developments including the minimal essential scheme and $N$-type cutoffs. The use of the Wetterich equation in explicit computations is illustrated within the Einstein-Hilbert truncation which constitutes the simplest non-perturbative approximation of the gravitational renormalization group flow. We conclude with a brief summary and comments on recent developments originating from other quantum gravity programs.}
	
	\section*{Keywords} 
	Quantum gravity, asymptotic safety, renormalization group, Wetterich equation, Reuter fixed point, Einstein-Hilbert truncation, phase diagram
	\section{Introduction}
	\label{sec.intro}
	\setcounter{footnote}{0}
	Our theoretical understanding of nature rests on two pillars. The electroweak and strong force and their interactions with the elementary particles are described by the standard model of particle physics. This theory is formulated as a relativistic quantum field theory in Minkowski space. The description of gravity is provided by general relativity, a classical field theory which encodes the gravitational interactions in the dynamics of spacetime. Conceptually, these theories are on very different footing and 
    the construction of a framework unifying gravity with the laws of quantum mechanics is
    one of the key open questions in theoretical high-energy physics to date.
    
     An important insight along these lines is that the quantization techniques successful in the case of the standard model of particle physics do not extend to gravity in a straightforward way: the perturbative quantization of general relativity leads to a perturbatively non-renormalizable quantum field theory with new infinities appearing at every order in perturbation theory \cite{'tHooft:1974bx,Goroff:1985sz,Goroff:1985th,vandeVen:1991gw}. This has led to the advance of several physics principles which deviate from the principles of continuum quantum field theory in more or less radical ways, see \cite{Armas:2021yut,Loll:2022ibq} for recent non-technical accounts.

	The gravitational asymptotic safety program is one particular line of quantum gravity research. The program is conservative in the sense that it strives for a consistent and predictive theory of the gravitational interactions within the framework of quantum field theory by seeking a non-perturbative high-energy completion. Its core assumptions are that the gravitational degrees of freedom are encoded in the spacetime metric also at trans-Planckian scales. Moreover, the theory retains invariance under coordinate transformations.\footnote{This assumption distinguishes the gravitational asymptotic safety program from Ho\v{r}ava-Lifshitz gravity \cite{Horava:2009uw} where this symmetry requirement is reduced to foliation-preserving diffeomorphisms, see \cite{Rechenberger:2012dt} for a pedagogical discussion.} The asymptotic safety hypothesis then stipulates that
	\begin{enumerate}
		\item these ingredients give rise to an interacting renormalization group fixed point -- called the Reuter fixed point.
		\item this fixed point controls the gravitational dynamics at trans-Planckian scales.
	\end{enumerate}
From a phenomenological perspective one also requires that the renormalization group flow emanating from the Reuter fixed point connects to a low-energy regime where the dynamics matches the one of general relativity to a good approximation.

We stress that the central element of the gravitational asymptotic safety program -- the existence of the Reuter fixed point coming with suitable properties -- is not an input. It must be established based on first-principle computations. At the technical level, this requires tools applicable to quantum field theory beyond the realm of perturbation theory. This is a highly non-trivial endeavor. It took about 20 years from Weinberg's first formulation of the asymptotic safety hypothesis \cite{Weinberg:1976xy,Weinberg:1980gg} to the advent of renormalization group techniques which could be used to investigate this hypothesis in a systematic way \cite{Reuter:1996cp}. 

Nowadays, there are two complementary computational approaches which naturally lend themselves to the exploration of the asymptotic safety mechanism in the context of gravity. Causal Dynamical Triangulations \cite{Ambjorn:2012jv,Loll:2019rdj} and Euclidean Dynamical Triangulations \cite{Ambjorn:2013eha,Coumbe:2014nea,Rindlisbacher:2015ewa,Bassler:2021pzt,Asaduzzaman:2022kxz} use Monte Carlo techniques to investigate the phase space of quantum geometries resulting from the gravitational path integral. In this setting, the Reuter fixed point may manifest itself as a second-order phase transition \cite{Ambjorn:2011cg} which allows to take the continuum limit in a controlled way. Alternatively, the Reuter fixed point can manifest itself in (approximate) solutions of the Wetterich equation \cite{Wetterich:1992yh,Morris:1993qb}.
	
This chapter will provide a basic introduction to the ideas underlying the gravitational asymptotic safety program (Sec.\ \ref{sec.ea}) before introducing the Wetterich equation \cite{Wetterich:1992yh,Morris:1993qb} and its adaptation to gravity \cite{Reuter:1996cp} as one of the main computational tools in the program (Sec.\ \ref{sec.frge}). Sec.\ \ref{sec.eh} illustrates how this tool is used in practical computations by working out the example of the Einstein-Hilbert truncation in a modern, background-independent way. Sec.\ \ref{sec.conclusion} provides our conclusion and a brief comments on renormalization group techniques implemented by other approaches to quantum gravity.

We stress that the exposition in this chapter is necessarily incomplete since it seeks to provide a concise introduction to the gravitational asymptotic safety program and the functional renormalization group which is accessible to a broader quantum gravity audience. For further details the reader is invited to consult the text books \cite{Percacci:2017fkn,Reuter:2019byg}, lecture notes \cite{Nagy:2012ef,Reichert:2020mja}, and general reviews \cite{Niedermaier:2006wt,Codello:2008vh,Reuter:2012id}. General introductions to the functional renormalization group are provided in \cite{Berges:2000ew,Gies:2006wv,Pawlowski:2005xe,Dupuis:2020fhh} and there are topical reviews focusing on asymptotic safety in the presence of matter fields \cite{Eichhorn:2018yfc}, the fluctuation approach to asymptotic safety \cite{Pawlowski:2020qer}, and its applications in the context of black holes \cite{Koch:2014cqa} and cosmology \cite{Bonanno:2017pkg}. Open issues have been discussed in the community report \cite{Bonanno:2020bil}.
	\section{The Asymptotic Safety Mechanism}
		\label{sec.ea}
	The insight that gravity could be asymptotically safe dates back to the seminal work of Weinberg \cite{Weinberg:1976xy,Weinberg:1980gg}. This initial proposal advocated asymptotic safety as a mechanism which renders physical scattering amplitudes finite (but non-vanishing) at energy scales exceeding the Planck scale. Motivated by computations  showing that gravity in $d=2+\epsilon$ spacetime dimensions possesses a non-trivial renormalization group (RG) fixed point \cite{Gastmans:1977ad,Christensen:1978sc}, it was suggested that this family of fixed points admits an analytic continuation up to $d=4$ where the corresponding fixed point should provide the high-energy completion of the gravitational interactions. The link between scattering amplitudes being finite and the RG fixed point builds on the insight that at such a fixed point all dimensionless quantities remain finite. If the fixed point controls the high-energy behavior, this property will also carry over to scattering amplitudes, which by themselves are dimensionless objects. This heuristic argument implies that it is not necessary that all dimensionless couplings remain finite. It suffices that the subset of couplings entering into physical observables (called essential couplings) attain their fixed-point values, as this is sufficient to ensure that the observables are well-behaved. A more detailed analysis of this scenario within the amplitude approach to asymptotic safety \cite{Draper:2020bop,Draper:2020knh,Knorr:2021iwv} revealed that there must be intricate relations between couplings and propagators. Most likely, these arise  as a consequence of quantum scale symmetry realized at the fixed point \cite{Wetterich:2019qzx}.
	
	The starting point for developing the idea of Asymptotic Safety is the functional integral over all Euclidean metrics,
	\be\label{eq.Zdef}
	Z = \int \cD h \, e^{-S[h]} \, ,
	\ee
	which would allow to determine all physical quantities of interest. In this respect, Asymptotic Safety shares the same starting point as Monte Carlo approaches to quantum gravity, foremost the Causal Dynamical Triangulation \cite{Ambjorn:2012jv,Loll:2019rdj} and Euclidean Dynamical Triagulation \cite{Ambjorn:2013eha,Coumbe:2014nea,Rindlisbacher:2015ewa,Bassler:2021pzt,Asaduzzaman:2022kxz} programs as well as Quantum Regge Calculus \cite{Rocek:1981ama,Hamber:2009mt}. 
	
	The functional renormalization group then recasts the problem of performing this functional integral into the problem of solving a functional differential equation, the Wetterich equation for the effective average action $\Gamma_k$ \cite{Wetterich:1992yh,Morris:1993qb,Reuter:1993kw,Reuter:1996cp} (derived in Sec.\ \ref{sec.frge}):
	\be\label{eq.Wetterich}
	k \p_k \Gamma_k = \frac{1}{2} {\rm Tr}\left[ \left( \Gamma^{(2)}_k + \cR_k \right)^{-1} k \p_k \cR_k \right] \, . 
	\ee
	Here $k$ is the coarse-graining scale and the trace contains an integration over loop momenta. The Wetterich equation implements the Wilsonian picture of renormalization in the following way: The  regulator $\cR_k$ appearing on the right-hand side separates the fluctuations into low- and high-momentum modes with respect to $k$. The change of $\Gamma_k$ is then governed by integrating out quantum fluctuations with momenta $p^2 \approx k^2$. In this way, one arrives at a formulation that is much better behaved as the initial problem of solving the functional integral \eqref{eq.Zdef} in one stroke.
	
	By construction, the propagators and vertices in the effective average action $\Gamma_k$ include the quantum corrections due to the high-momentum fluctuations. In this sense, it provides an effective description of physics at length scales $l \sim k^{-1}$. This makes $\Gamma_k$ a quite complicated object. Its natural habitat is the theory space $\cT$. By definition, this space consists of all action functionals $A[\cdot]$ which can be constructed from the field content of the theory and meets its symmetry requirements. In the context of gravity, where the field content is given by (Euclidean) spacetime metrics $g_{\mu\nu}$, prototypical examples for these building blocks include the terms appearing in the Einstein-Hilbert action,
	\be\label{Oexamples}
	 \cO_1 = \int d^dx \sqrt{g} \, , \qquad \cO_2 = \int d^dx \sqrt{g} R \, , 
	 \ee
	 where $\sqrt{g} \equiv \sqrt{\det(g)}$ and $R$ is the Ricci scalar constructed from $g_{\mu\nu}$ (also see Table \ref{tab.derivativeexp} for further examples). Given a basis $\{\cO_i\}$ for these monomials, the effective average action can be expanded in this basis
	 \be\label{eq.expansionG}
	 \Gamma_k = \sum_i \, \bar{u}^i(k) \, \cO_i \, .
	 \ee
	 The dependence on the coarse-graining scale is captured by the dimensionful couplings $\bar{u}^i(k)$. For the purpose of studying RG flows it is useful to trade these dimensionful couplings with their dimensionless counterparts obtained by rescaling with $k$,
	 \be\label{eq.dimless}
	 u^i(k) \equiv \bar{u}^i(k) \, k^{-d_i} \, , 
	 \ee
	 where $d_i \equiv [\bar{u}^i]$ is the mass-dimension of the coupling. The couplings $u^i$ then serve as coordinates on $\cT$.
	 
	 Evaluating \eqref{eq.Wetterich} for the expansion \eqref{eq.expansionG} gives the component form of the functional renormalization group equation
	 \be\label{def.betafcts}
	 k \p_k \, u^i(k) = \beta^i(\{u^j\}) \, . 
	 \ee
	 The beta functions $\beta^i(\{u^j\})$ capture the dependence of the dimensionless couplings on the coarse-graining scale. Dimensional analysis entails that the functions $\beta^i(\{u^j\})$ are independent of $k$, since this is the only dimensionful object in the construction. Thus Eq.\ \eqref{def.betafcts} constitutes an infinite-dimensional system of coupled, autonomous, first order differential equations. Its solutions are called RG trajectories. The problem of performing the functional integral \eqref{eq.Zdef} is then translated into finding globally well-defined RG trajectories
	 \be\label{def.rgtraject}
	 k \rightarrow \Gamma_k \, , \qquad k \in [0,\infty] \, , 
	 \ee
	 which exist for all values of the coarse-graining scale $k$.
	 \begin{figure}[t!]
	 	\centering
	 	\includegraphics[width=.9\textwidth]{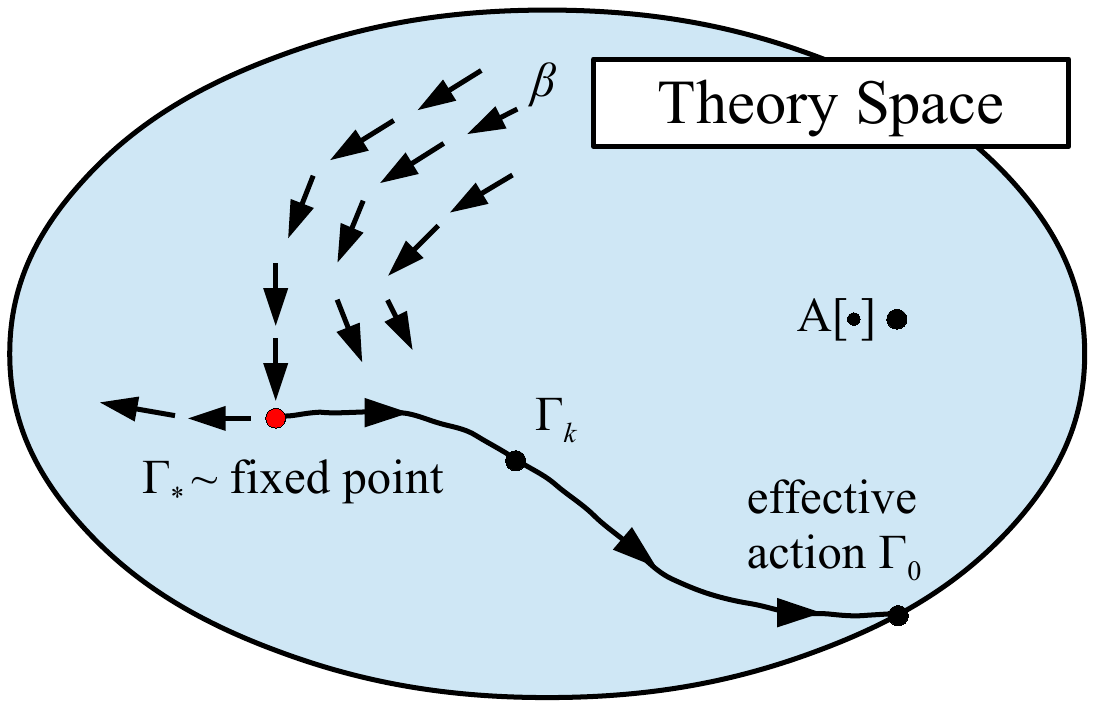}
	 	\caption{\label{fig:theoryspace} Illustration of theory space and its structures: by definition, the theory space contains all action functionals $A[\cdot]$ which can be constructed from a given field content and obey the desired symmetries. The theory space comes with a vector field, the beta functions $\beta$. The integral curves of this vector field (RG trajectories) are exemplified by the black solid curve. The example emanates from a fixed point (red dot) with one UV-attractive (Re$\theta_I > 0$) and one UV-repulsive (Re$\theta_I < 0$) eigendirection. The endpoint of the RG trajectory at $k=0$ coincides with the effective action $\Gamma$. Conventionally, all arrows point towards a lower coarse-graining scale, i.e., in the direction of integrating out fluctuation modes.}
	 \end{figure}
	 
	 By definition, RG fixed points $\{u^j_*\}$ are stationary points of the system \eqref{def.betafcts}, satisfying
	 \be\label{eq.fpcond}
	 \beta^i(\{u^j_*\}) = 0 \, , \qquad \forall \, i \, . 
	 \ee
	 As a consequence, it takes infinite amount of ``RG-time'' for an RG trajectory to actually reach the fixed point. In this way fixed points can provide a well-defined limit $k\rightarrow \infty$ in which all dimensionless couplings $u^i(k) \rightarrow u^i_*$ remain finite. Thus, fixed points are natural candidates for providing a well-defined high-energy completion of a theory. It is this concept that underlies the Wilsonian picture of renormalization.
	 
	 At this point it is interesting to inquire about the conditions for an RG trajectory being dragged into a fixed point as $k \rightarrow \infty$. This question is closely related to the predictive power of the construction. In the vicinity of a fixed point $\{u^i_*\}$, the properties of the RG flow can be studied by linearizing the system \eqref{def.betafcts},
	 \be\label{linflow}
	 k \p_k u^i(k) = \sum_j B^i{}_j \, \left(u^j(k) - u_*^j \right) + O(u^2) \, . 
	 \ee
	 Here
	 \be\label{def.B}
	 B^i{}_j \equiv \left. \frac{\p}{\p u^j} \beta^i \right|_{u = u^*}
	 \ee
	 is the stability matrix associated with the fixed point. The solutions of \eqref{linflow} are readily given in terms of the right-eigenvectors $V_I$ and stability coefficients $\theta_I$ of $B$,
	 \be\label{def.theta}
	 \sum_j B^i{}_j \, V^j_I = - \theta_I \, V_I^i \, , \qquad \forall \, I \, , 
	 \ee
	 and take the form
	 \be\label{sollin}
	 u^i(k) = u^i_* + \sum_J C_J \, V_J^i \left( \frac{k_0}{k} \right)^{\theta_J} \, . 
	 \ee
	 Here $C_J$ are constants of integration and $k_0$ denotes an arbitrary reference scale.
	 
  Inspecting \eqref{sollin} reveals that eigendirections with Re$(\theta_I) > 0$ are attracted by the fixed point as $k \rightarrow \infty$ while the ones with Re$(\theta_I)<0$ are repulsive in this limit. The corresponding scaling operators are called ``UV-relevant'' and ``UV-irrelevant'', respectively. This suggests splitting the set $\{C_I\}$ according to
  \be\label{Crel}
  \{C_I^{\rm relevant}\} = \{ C_I \, | \, {\rm Re}(\theta_I) > 0 \} \, , \qquad \{C_I^{\rm irrelevant}\} = \{ C_I \, | \, {\rm Re}(\theta_I) < 0 \} \, . 
  \ee
  The case Re$\theta_I$=0 corresponds to a marginal direction. Determining whether this direction is UV-attractive or UV-repulsive requires going beyond the linear approximation \eqref{sollin} and will not be discussed in detail here.
  
  The condition that the fixed point controls the UV-behavior of the RG-trajectory then enforces $C_I^{\rm irrelevant} = 0$, for all $I$. The solutions meeting this condition span the UV-critical hypersurface of the fixed point. The $\{C_I^{\rm relevant}\}$ are the free parameters of the construction and label the solutions within this hypersurface. Their value is unconstrained by demanding a well-defined UV-completion and must be determined by other theoretical considerations or experimental input. This discussion also shows that fixed points with a lower-dimensional UV-critical hypersurface have a higher predictive power.
  
  Up to this point, our discussion of a high-energy completion referred to a generic renormalization fixed point. It is then customary to distinguish among a Gaussian fixed point (GFP) and a non-Gaussian fixed point (NGFP). The definition of the former is that the critical exponents of its stability matrix agree with the canonical mass-dimension of the corresponding coupling $\theta_I = d_I$. This signals that the underlying theory is the free theory. At a NGFP, the stability coefficients receive quantum corrections,
  \be\label{theta.NGFP}
  \theta_I = d_I + \text{quantum corrections} \, . 
  \ee
  The latter indicate that the theory linked to the fixed point is interacting. Notably, this definition of a Gaussian and non-Gaussian fixed point is not based on the values $\{u_*^i\}$. Since the spectrum of the stability matrix is invariant under a redefinition $u^i \mapsto \tilde{u}^i(\{u^j\})$, this characterization is independent of a specific choice of ``coordinate system'' on $\cT$. An important subset of NGFPs are ``almost-Gaussian'' NGFPs. In this case the quantum corrections in \eqref{theta.NGFP} are weak in the sense that the $\theta_I$ are dominated by their classical part. This implies that classical power-counting is still a valid guiding principle for determining whether a scaling operator is relevant or irrelevant. Beyond the class of ``almost Gaussian'' NGFPs, there could also be fixed points where the critical exponents are dominated by quantum effects. The systematic investigation of this possibility is beyond the scope of most of current searches for RG fixed points based on functional renormalization group equations though.\footnote{Some insights on potential stability patterns associated with such fixed points have recently be discussed based on the composite operator equation \cite{Houthoff:2020zqy,Kurov:2020csd}, indicating that studying such fixed points requires approximations at a significant level of complexity as well as dedicated search strategies.} Depending on whether the high-energy completion is provided by a GFP or a NGFP, the theory is termed ``asymptotically free'' or ``asymptotically safe''. A prototypical example of the former case is Quantum Chromodynamics while the latter case is realized by gravity in $d=2+\epsilon$ spacetime dimensions \cite{Gastmans:1977ad,Christensen:1978sc}.
  
  We conclude this section with two clarifications. For a globally well-defined RG trajectory, the solutions \eqref{def.rgtraject} interpolate between the microscopic dynamics determined by the RG fixed point for $k\rightarrow\infty$ and the standard effective action $\lim_{k \rightarrow 0} \Gamma_k = \Gamma$. \emph{All physics should then be extracted from $\Gamma$ using its quantum corrected propagators and vertices.} Similarly to \eqref{eq.expansionG}, $\Gamma$ can be expanded in a basis of the theory space
  \be\label{EA.exp}
  \Gamma = \sum_i \, \bar{u}^i_{\rm eff} \, \cO_i \, ,
  \ee
  with the relation between the couplings being $\bar{u}^i_{\rm eff} = \lim_{k \rightarrow 0} \bar{u}^i(k)$. This expansion is similar to the one encountered in effective field theory where the $\cO_i$ are organized according to their canonical mass-dimension and the sum is truncated at a given order. The key difference to the effective field theory approach is that the RG flow determines the effective couplings in terms of the free coefficients $\{C_I^{\rm relevant}\}$:
  \be\label{predictions}
  \bar{u}^i_{\rm eff} = \bar{u}^i_{\rm eff}(C_I^{\rm relevant}) \, . 
  \ee
  Provided that there are more couplings $\bar{u}^i_{\rm eff}$ than free parameters $C_I^{\rm relevant}$, the high-energy completion induces a (potentially infinite number of) relations between the effective couplings. These provide predictions which can be confronted with theoretical consistency requirements and experimental data. On this basis one can deduce whether a given RG fixed point leads to low-energy physics compatible with nature. This also allows to falsify the construction, provided that the properties of the fixed point and its UV-critical surface are known at a sufficient level of detail. 
  
  We also stress that the dependence of couplings on the coarse-graining scale $k$ should not be identified with the running of a coupling with respect to a physical energy scale, see \cite{Donoghue:2019clr,Bonanno:2020bil} for instructive examples. Generically, the couplings appearing in \eqref{predictions} are not constant but come in the form of form factors depending on the momenta of the fields in a non-trivial way. In the simplest case (cf.\ \eqref{eq.formfactors}) this dependence contains a single momentum scale
  \be
  \bar{u}^i_{\rm eff} \rightarrow \bar{u}^i_{\rm eff}(p^2) \, .
  \ee 
  In practice, the value of the coupling is then measured at a fixed momentum scale $\mu^2$. The non-trivial $p$-dependence then induces the ``running'' of the coupling with respect to its value determined at the reference scale. In this simplest case, this is the logarithmic running of a dimensionless coupling seen in perturbation theory, but the momentum dependence can be significantly more involved than that.
	\section{The Functional Renormalization Group}
	\label{sec.frge}
	The basic idea of a functional renormalization group equation (FRGE) is to recast the functional integral over quantum fluctuations in terms of a functional differential equation. The FRGE implements Wilson's modern viewpoint on renormalization \cite{Wilson:1973jj}: in contrast to a perturbative approach based on evaluating Feynman diagrams, quantum fluctuations are not integrated over in one stroke. Instead they are integrated out ``shell-by-shell'' in momentum space starting with the most energetic ones. This leads to a one-parameter family of effective actions $\Gamma_k$ whose propagators and vertices already contain the quantum corrections from fluctuations with momenta $p^2 \gtrsim k^2$. The textbook effective action $\Gamma$ is recovered in the limit where all fluctuations are integrated out, $\Gamma = \lim_{k \rightarrow 0} \Gamma_k$. 
	
	The FRGE most frequently used in hands-on computations is the Wetterich equation \cite{Wetterich:1992yh,Morris:1993qb,Reuter:1996ub,Reuter:1996cp}. This section reviews its construction for scalar fields (Sec.\ \ref{sect.31}) before extending the formalism to gravity (Sec.\ \ref{sect.32}). The most common non-perturbative approximation techniques to this equation are introduced  in Sec.\ \ref{sect.34} and important extensions giving structural insights to the gravitational renormalization group flow are summarized in Sec.\ \ref{sect.33}.
	\subsection{The Wetterich equation for scalar field theory}
	\label{sect.31}
	The Wetterich equation is a universal tool for studying the RG flow of theories built from essentially any field content \cite{Dupuis:2020fhh}. In order to introduce this tool with the absolute minimum of technicalities, we first focus on a real scalar field $\varphi$ living on a $d$-dimensional Euclidean spacetime $(\mathbb{R}^d, \delta_{\mu\nu})$. For pedagogical reasons, we first review the construction of the effective action in this setting before introducing the effective average action and its FRGE.

	We start from the generating functional of correlation functions (path integral)
	\be\label{defZ}
	Z[J] \equiv \frac{1}{N} \int \cD\varphi \, \exp\left\{ -S[\varphi] + \int d^dx \, J(x) \varphi(x) \right\} \, . 
	\ee
	Here $N \equiv \int \cD\varphi \, \exp\left\{ -S[\varphi]\right\}$ is a normalization factor and $J(x)$ a source coupling to the quantum field. The dynamics of the field is governed by the bare action $S[\varphi]$ which is kept arbitrary at this point. Generically, this generating functional diverges and we implicitly assume that it has been suitably regularized by including an UV-cutoff. Eq.\ \eqref{defZ} allows to construct expectation values of operators $\cO$
	\be\label{defnpt}
	\left\langle \cO[\varphi] \right\rangle \equiv  \frac{1}{N} \int \cD\varphi \, \cO[\varphi] \, \exp\left\{ -S[\varphi] \right\} \, . 
	\ee
	In particular, expectation values of operators polynomial in $\varphi$ can be obtained  by taking functional derivatives with respect to the source and subsequently setting $J$ to zero
	\be\label{defcorrelators}
	\langle \varphi(x_1) \cdots \varphi(x_n) \rangle = \left.  \frac{\delta^n Z[J]}{\delta J(x_1) \cdots \delta J(x_n)} \right|_{J=0} \, . 
	\ee
	Here, the normalization factors are chosen such that $\langle \unit \rangle = 1$. Based on the path integral \eqref{defZ}, one obtains the functional $W[J]$ generating all connected Green's functions by setting
	\be\label{defW}
	Z[J] \equiv e^{W[J]} \, . 
	\ee
	We then introduce the mean field $\phi(x)$ as the expectation value of $\varphi(x)$:
	\be\label{defmean}
	\phi(x) = \langle \varphi(x) \rangle = \frac{\delta W[J]}{\delta J(x)} \, . 
	\ee 
	Finally, one constructs the effective action $\Gamma[\phi]$ as the Legendre transform of $W[J]$. If the relation \eqref{defmean} can be solved for the source, giving $J[\phi]$, it takes the form\footnote{In the general case, the effective action is obtained as the Legendre-Fenchel transform $\Gamma[\phi] = \sup_{J(x)}\left( \int d^dx \, J[\phi](x) \phi(x) - W[J[\phi]] \right)$. In the sequel, formulas are understood to include the supremum if needed.} 
	\be\label{defeffectiveaction}
	\Gamma[\phi] = \int d^dx \, J[\phi](x) \phi(x) - W[J[\phi]] \, . 
	\ee
	The fact that $W[J]$ and $\Gamma[\phi]$ are related by a Legendre transform implies that
	\be\label{prop.inverse}
	\int d^dy \, \frac{\delta^2 W[J]}{\delta J(x_1) \delta J(y)} \, \frac{\delta^2 \Gamma[\phi]}{\delta \phi(y) \delta \phi(x_2)} = \delta^d(x_1 - x_2) \,  .
	\ee
	
	The effective action provides the equation of motion for the mean field in the presence of a source,
	\be
	\frac{\delta \Gamma[\phi]}{\delta \phi(x)} = J(x) \, . 
	\ee
	Higher order functional derivatives generate the one-particle irreducible ($1$PI) $n$-point functions
	\be
	\Gamma^{(n)}[\phi] \equiv \frac{\delta^n \Gamma[\phi]}{\delta \phi(x_1) \cdots \delta\phi(x_n)} = \langle \varphi(x_1) \cdots \varphi(x_n) \rangle_{1 {\rm PI}} \, . 
	\ee
	Eq.\ \eqref{prop.inverse} then entails that the second functional derivative of $\Gamma[\phi]$ encodes the quantum corrected propagator
	\be
	\left( \Gamma^{(2)}(x_1,x_2) \right)^{-1} = W^{(2)}(x_1,x_2) = G(x_1,x_2) \, . 
	\ee
	Scattering processes are described by tree-level Feynman diagrams constructed from the propagators and vertices extracted from $\Gamma[\phi]$. In this sense, the effective action is the quantum analog of the classical action, since it encodes the quantum physics at tree level. Determining $\Gamma[\phi]$ is therefore often considered as equivalent to solving the quantum theory.
	
	The construction of the effective average action $\Gamma_k[\phi]$ proceeds along very similar lines. The key modification occurs at the level of the generating functional \eqref{defZ} which is supplemented by an IR-regulator
	\be\label{defreg}
	\Delta S_k[\varphi] = \frac{1}{2} \int d^dx \, \varphi(x) R_k(-\p^2) \varphi(x) \, . 
	\ee
	The purpose of this extra ingredient is to provide a $k$-dependent mass-term for quantum fluctuations with moments $p^2 \ll k^2$. In the simplest case, this is implemented by requiring that the regulator $R_k(p^2)$ satisfies
	\be\label{reg.prop}
	R_k(p^2) \approx
	\left\{ 
	\begin{array}{ll}
		k^2 & \quad \text{for} \; p^2 \ll k^2 \, , \\
		0 & \quad \text{for} \; p^2 \gg k^2 \, . 
	\end{array}
	\right.
	\ee
	Examples of regulators used in practical computations include the (smooth) exponential cutoff,
	\be\label{Rexp}
	R_k(p^2) = p^2 \left( \exp(p^2/k^2) - 1 \right)^{-1} \, , 
	\ee
	and Litim-type regulators,
	\be\label{Rlitim}
	R_k(p^2) = (k^2 - p^2) \Theta(1-p^2/k^2) \, , 
	\ee
	where $\Theta(x)$ is the Heaviside step function. Adding \eqref{defreg} to the weight in the generating functional \eqref{defZ} induces a dependence on the scale $k$
	\be\label{defZk}
		Z_k[J] = \frac{1}{N} \int \cD\varphi \, \exp\left\{ -S[\varphi] - \Delta S_k[\varphi] + \int d^dx \, J(x) \varphi(x) \right\} \, .
	\ee
	The effect is that the contribution of modes with $p^2 \ll k^2$ to the generating functional becomes suppressed while the modes with $p^2 \gg k^2$ are integrated out in the usual way. Thus $k$ acquires a natural interpretation as a coarse-graining scale, marking the scale up to which microscopic quantum fluctuations are included in the generating functional.
	
	Following the steps leading to the effective action, we then define the (now $k$-dependent) generating functional for connected Green's functions $W_k[J]$ by
	\be\label{defWk}
	Z_k[J] = \exp[W_k[J]] \, . 
	\ee
	By definition, the effective average action is then given by a modified Legendre transform of $W_k[J]$:
	\be\label{def.eaa}
	\Gamma_k[\phi] \equiv \int d^dx \, J[\phi](x) \phi(x) - W_k[J] - 	\Delta S_k[\phi] \, . 
	\ee
	For $k=0$ the IR regulator in the definition of $W_k[J]$ as well as $\Delta S_k[\phi]$ vanish and \eqref{def.eaa} agrees with the definition of the effective action \eqref{defeffectiveaction}:
	\be
	\lim_{k \rightarrow 0} \Gamma_k[\phi] = \Gamma[\phi] \, . 
	\ee
	
	The key virtue of the effective average action is that its $k$-dependence is governed by a functional renormalization group equation, the Wetterich equation. This equation is formally exact in the sense that no approximations are made in its derivation. The construction of the Wetterich equation then proceeds along the following lines. We start by introducing the RG time $t \equiv \ln k/k_0$, with $k_0$ being an arbitrary reference scale, so that $\p_t = k \p_k$. We then consider the auxiliary generating functional
	\be
	\tilde{\Gamma}_k[\phi] \equiv \int d^dx \, J[\phi](x) \, \phi(x) - W_k[J] \, . 
	\ee
	Taking a partial derivative of this definition with respect to RG time yields
	\be\label{ptgammatilde}
	\begin{split}
	\p_t \tilde{\Gamma}_k[\phi] = & - \p_t W_k[J]= \frac{1}{2} \int d^dx \int d^dy \; \langle \varphi(x) \varphi(y) \rangle \; \p_t R_k(x,y) \, .
	\end{split}
\ee
	Here we have used that $\p_t W_k[J] = \p_t \ln Z_k[J]$ with 
	\be
	\begin{split}
		 \p_t \ln Z_k[J] 
		& \, = - \frac{1}{2 Z_k} \int \cD \varphi \int d^dx \int d^dy \, \varphi(x) \, \p_t R_k(x,y) \, \varphi(y) \, \times \\ & \qquad \times \exp\left\{ -S[\varphi] - \Delta S_k[\varphi] + \int d^dx \, J(x) \varphi(x) \right\} \\
		& \, = - \frac{1}{2}  \int d^dx \int d^dy \,  \,  \langle \varphi(x) \varphi(y) \rangle  \,  \p_t R_k(x,y) \,  \, ,   
	\end{split}
    \ee
    in the second step.    We then introduce the ($k$-dependent) mean field
    \be
    \phi(x) = \langle \varphi(x) \rangle = \frac{\delta W_k[J]}{\delta J(x)} \, , 
    \ee
    together with the two-point functions
    \be
    \langle \varphi(x) \varphi(y) \rangle_{\rm c} \equiv \frac{\delta^2 W_k[J]}{\delta J(x) \delta J(y)} 
    \, , \qquad \tilde{\Gamma}_k^{(2)}(x,y) \equiv \frac{\delta^2 \tilde{\Gamma}_k[\phi]}{\delta \phi(x) \delta \phi(y)} \, . 
    \ee
    Since $\tilde{\Gamma}_k[\phi]$ and $W_k[J]$ are again related by a Legendre transform, these functionals are again each others inverse, cf.\ Eq.\ \eqref{prop.inverse}. This allows to express the two-point function appearing in the relation \eqref{ptgammatilde} in terms of $\tilde{\Gamma}_k^{(2)}(x,y)$
    \be
    \begin{split}
    \langle \varphi(x) \varphi(y) \rangle = & \, \langle \varphi(x) \varphi(y) \rangle_{\rm c} + \langle \varphi(x) \rangle \, \langle \varphi(y) \rangle \, \\
    = & \left( \tilde{\Gamma}_k^{(2)}(x,y) \right)^{-1} + \phi(x) \phi(y) \, .
    \end{split}
    \ee
    Here we used the definition of the (now $k$-dependent) mean field when recasting the last term. Substituting this relation into \eqref{ptgammatilde} then yields
    \be
    \p_t \tilde{\Gamma}_k = \frac{1}{2} \int d^dx \int d^dy \, \left[ \left( \tilde{\Gamma}_k^{(2)}(x,y) \right)^{-1} \, \p_t R_k(x,y) \right] + \p_t \Delta S_k[\phi] \, . 
    \ee
    Bringing the second term to the left-hand side and using that $\Gamma_k = \tilde{\Gamma}_k - \Delta S_k$ allows to rewrite this equation in terms of the effective average action
    \be\label{FRGEint}
    \p_t \Gamma_k[\phi] = \frac{1}{2} \int d^dx \int d^dy \, \left[ \, \left( \Gamma_k^{(2)} + R_k \right)^{-1} \, \p_t R_k \, \right] \, . 
    \ee
    Noticing that the integrals on the right-hand side actually correspond to taking the trace of the argument, we arrive at the Wetterich equation in its iconic form
	\be\label{FRGE}
	\p_t \Gamma_k = \frac{1}{2} {\rm Tr}\left[ \left( \Gamma_k^{(2)} + R_k \right)^{-1} \, \p_t R_k \right] \, . 
	\ee
	
	The Wetterich equation exhibits several remarkable features arising from the interplay of  $R_k(p^2)$ in the numerator and denominator of the trace argument. In the propagator term $\left( \Gamma_k^{(2)} + R_k \right)^{-1}$, the regulator provides a mass to the fluctuations, ensuring the absence of IR-singularities as long as $k$ is finite. In the numerator, the condition $R_k(p^2) \rightarrow 0$ for $p^2 \gg k^2$ entails that the trace argument vanishes for high-momentum modes. As a consequence the right-hand side is IR and UV-finite and any UV-regulator implicit in the definition of the initial functional integral can be removed trivially.\footnote{In some practical computations, as e.g.\ in the computation of spectral flows \cite{Braun:2022mgx}, one may want to resort to regulators $R_k(p^2)$ where this fall-off property in the UV does not hold. In this case, the flow equation must be supplemented by additional counterterms absorbing the UV-divergences.}
	
	The regulator structure furthermore entails that the trace argument is peaked at momenta $p^2 \approx k^2$. Hence the flow of $\Gamma_k[\phi]$ is driven by integrating out quantum fluctuations whose momenta are comparable to the coarse-graining scale $k$. In this way the Wetterich equation implements the Wilsonian picture of renormalization, integrating out quantum fluctuations shell-by-shell in momentum space. Notably, Eq.\ \eqref{FRGE} allows to start from any initial condition $\Gamma_\Lambda$ and integrate its RG flow towards the infrared. Thus the Wetterich equation does not require specifying a bare action a priori. These are obtained as the fixed points of the RG flow through the reconstruction problem \cite{Manrique:2008zw}.  
	
	We also observe that the combination of propagator and regulator within the trace induces a projective feature. Any $k$-independent rescaling of the fluctuation field affects the regulator and propagator in the same way, so that such rescalings drop out from the right-hand side of the equation. This renders the flow equation invariant with respect to certain classes of field redefinitions. 
	
	\subsection{The Wetterich equation for gravity}
	\label{sect.32}
In the previous section, we derived the Wetterich equation \eqref{FRGE} for a real scalar field. Its extension to gauge fields and fermions is conceptually straightforward. In the context of gravity the construction faces two conceptual obstacles though. Firstly, our understanding of classical gravity based on general relativity indicates that gravitational interactions are mediated through the curvature of spacetime. This implies that spacetime itself becomes a dynamical and, in the context of the quantum theory, also fluctuating object. Hence, the concept of a fixed, non-dynamical spacetime providing the stage for the dynamics is lost at this point. This raises the question about how to define the coarse-graining scale $k$. Secondly, gravity shares some properties of a gauge theory. The Einstein-Hilbert action, for example, is invariant under coordinate transformations which act on the metric according to
\be\label{eq.coordtrafo}
\delta g_{\mu\nu} \equiv \cL_v g_{\mu\nu} = v^\rho \p_\rho g_{\mu\nu} + (\p_\mu v^\rho) g_{\rho\nu} + (\p_\nu v^\rho) g_{\rho\mu} \, . 
\ee
Here $\cL_v$ denotes the Lie derivative along the generating vector field $v^\mu$. In order to ensure that the generating functional $Z_k$ sums over physically inequivalent configurations only, one has to introduce a suitable gauge-fixing condition. By construction, the gauge-fixing term breaks the invariance under the transformations \eqref{eq.coordtrafo}. As a consequence, the effective (average) action may loose this symmetry, leading to a proliferation of interaction monomials which could be generated along the RG flow.

Following the seminal work by Reuter \cite{Reuter:1996cp}, both of these conceptual difficulties can be overcome by resorting to the background field method. This procedure splits the (Euclidean) quantum metric $g_{\mu\nu}$ into a generic (but non-fluctuating) background metric $\gb_{\mu\nu}$ and fluctuations around this background $h_{\mu\nu}$. There is no requirement that the latter are small. The decomposition can then be implemented either through a linear or an exponential split (see \cite{Ohta:2016npm,Ohta:2016jvw} for a detailed discussion):
\be\label{eq.backgrounddecomp}
g_{\mu\nu} = \gb_{\mu\nu} + h_{\mu\nu} \, , \qquad g_{\mu\nu} = \gb_{\mu\alpha} \left( e^h \right)^\alpha{}_\nu \, . 
\ee
Here we follow the standard convention that indices are raised and lowered with the background metric, i.e., $h^\mu{}_\nu = \gb^{\mu\alpha} h_{\alpha\nu}$, etc. While these decompositions agree to leading order in the fluctuation field, they actually define different theories, since they do not cover the same space of quantum fluctuations. Heuristically, this can be argued based on the observation that the linear split allows for  $g_{\mu\nu}$ and $\gb_{\mu\nu}$ having different signatures while in the exponential split this is not the case \cite{Demmel:2015zfa}. This is also confirmed by computing properties of the Reuter fixed point in $d=2+\epsilon$ dimensions \cite{Nink:2015lmq}. In the following, we will adopt the linear split for simplicity.

The background metric then allows to quantize metric fluctuations along the lines of quantum field theory in a curved spacetime. Moreover, it allows to circumvent the conceptual difficulties discussed above as follows. Firstly, it provides the basis for separating fluctuations into ``high-'' and ``low-''momentum modes relative to the coarse-graining scale in a purely geometric way. Taking the background to be compact and introducing the Laplacian $\Delta \equiv - \gb^{\mu\nu} \Db_\mu \Db_\nu$ constructed from the background metric, one can obtain the ordered set of eigenmodes
\be\label{eigenmodes}
\Delta h^{n}_{\mu\nu} = E_n \, h^{n}_{\mu\nu} \, , \qquad n=0,1,\cdots \, , 
\ee
with $E_0 \le E_1 \le E_2 \le \cdots$. Fluctuations with $E_n \lesssim k^2$ are then considered ``long-range'' and are suppressed by the regulator while ``short-range'' fluctuations characterized by $E_n \gtrsim k^2$ are integrated out without suppression factor. Practically, this is achieved by generalizing \eqref{defreg} to
\be\label{defreggrav}
\Delta S_k[h;\gb] = \frac{1}{2} \int d^dx \sqrt{\gb} \left[ h_{\mu\nu}(x) \, \cR_k^{\mu\nu\alpha\beta}(\Delta) \, h_{\alpha\beta}(x) \right] \, . 
\ee
The switch from $R_k$ to $\cR_k$ anticipates that, in general, the regulator is a matrix in field space carrying a non-trivial tensor structure. Note that \eqref{defreggrav} is quadratic in the fluctuation field: specifically, $\cR_k^{\mu\nu\alpha\beta}(\Delta)$ is independent of the fluctuation field and depend on $\gb_{\mu\nu}$ only. This property is essential in order to arrive at a FRGE of the form \eqref{FRGE}.  

Secondly, the linear split allows to realize the transformation \eqref{eq.coordtrafo} in two distinct ways. Quantum gauge transformations ($Q$) keep $\gb_{\mu\nu}$ fixed and attribute the transformation of $g_{\mu\nu}$ to the fluctuation field
\be\label{def.Qtrafo}
\delta^Q \gb_{\mu\nu} = 0 \, , \qquad \delta^Q h_{\mu\nu} = \cL_v(\gb_{\mu\nu}+h_{\mu\nu}) \, . 
\ee
It is this transformation that must be gauge-fixed. In addition, one can define background gauge transformations ($\delta^B$) where each field transforms as a tensor of the corresponding rank
\be\label{eq.Btrafo}
\delta^B \gb_{\mu\nu} = \cL_v \gb_{\mu\nu} \, , \qquad \delta^B h_{\mu\nu} = \cL_v h_{\mu\nu} \, . 
\ee
This transformation can be maintained as an auxiliary symmetry by resorting to the class of background covariant gauges. Following the Faddeev-Popov procedure, the gauge-fixing is implemented by supplementing the gravitational action $S[g]$ by a gauge-fixing term
\be\label{eq.gf}
S^{\rm gf}[h;\gb] = \frac{1}{2\alpha} \int d^dx \sqrt{\gb} \, \gb^{\mu\nu} F_\mu \, F_\nu \, . 
\ee
Here, $\alpha$ is a free parameter and the gauge-fixing condition $F_\mu[h;\gb]$ transforms as a rank-one tensor with respect to \eqref{eq.Btrafo}.

The gauge-fixing term is accompanied by the action for the Faddeev-Popov ghost and anti-ghost fields $C^\mu$ and $\bar{C}_\mu$
\be\label{Sghost}
S^{\rm ghost}[h,\Cb,C;\gb] = - \sqrt{2} \int d^dx \sqrt{\gb} \, \Cb_\mu \, \gb^{\mu\nu} \, \frac{\delta F_\nu}{\delta h_{\alpha\beta}} \, \cL_C(\gb_{\alpha\beta} + h_{\alpha\beta}) \, .
\ee 
This action exponentiates the Faddeev-Popov determinant
\be
\det \cM = \det \left[ \frac{\delta F_\mu}{\delta v^\nu} \right] = \int \cD C^\mu \cD \bar{C}_\nu \, e^{- \int \Cb \cM C} \, . 
\ee

At this point we have all the ingredients to write down the analogue of the generating functional \eqref{defWk} in the context of gravity
\be\label{defZkgrav}
\begin{split}
\exp(W_k[J;\gb]) =  \frac{1}{N} \int  &\, \cD h_{\alpha\beta} \cD C^\mu \cD \bar{C}_\nu \, \exp\Big\{
-S[\gb+h] - S^{\rm gf}[h;\gb]  \\ 
&  
- S^{\rm ghost}[h,\Cb,C;\gb] - \Delta S_k[h,\Cb,C;\gb] + S^{\rm source}
 \Big\} \, . 
\end{split} 
\ee
Here $S[g]$ denotes a generic action built from the metric $g_{\mu\nu}$, invariant under \eqref{eq.coordtrafo}, $S^{\rm gf}[h;\gb]$ and $S^{\rm ghost}[h,\Cb,C;\gb]$ are the gauge-fixing and ghost actions given in Eqs.\ \eqref{eq.gf} and \eqref{Sghost}, and $\Delta S_k[h,\Cb,C;\gb]$ is the IR regulator \eqref{defreggrav} extended by a $k$-dependent mass term for the ghost fields. Finally, 
\be\label{Ssource}
S^{\rm source} = \int d^dx \sqrt{\gb} \left\{ t^{\mu\nu} h_{\mu\nu} + \bar{\sigma}_\mu C^\mu + \sigma^\mu \Cb_\mu \right\}
\ee
introduces sources for the quantum field, which we collectively label by $J \equiv (t^{\mu\nu}, \sigma^\mu, \bar{\sigma}_\mu)$.

The construction of the effective average action then proceeds analogously to the scalar case. Taking functional derivatives of $W_k[J;\gb]$ with respect to the sources gives the expectation values of the fluctuation fields
\be\label{def.flucexp}
\langle h_{\mu\nu} \rangle = \frac{1}{\sqrt{\gb}} \frac{\delta W_k}{\delta t^{\mu\nu}} \, , \quad 
\langle \Cb_\mu \rangle = \frac{1}{\sqrt{\gb}} \frac{\delta W_k}{\delta \sigma^\mu} \, , \quad 
\langle C^\mu \rangle = \frac{1}{\sqrt{\gb}} \frac{\delta W_k}{\bar{\sigma}_\mu} \, . 
\ee
In a slight abuse of notation we then use the same labels for the mean- and quantum fields, identifying
\be\label{notationsimple}
h_{\mu\nu} = \langle h_{\mu\nu} \rangle \, , \qquad 
C^\mu = \langle C^\mu \rangle  \, , \qquad 
 \Cb_\mu  = \langle \Cb_\mu \rangle  \, , \qquad g_{\mu\nu} = \langle \gb_{\mu\nu} + h_{\mu\nu} \rangle \, . 
\ee

We then assume again that the field-source relations \eqref{def.flucexp} can be solved for the sources as functions of the mean field. The effective average action is then again defined as the modified Legendre transform of $W_k$:
\be\label{def.eaagrav}
\Gamma_k[\Phi;\gb] = \int d^dx \sqrt{\gb}  \left\{ t^{\mu\nu} h_{\mu\nu} + \bar{\sigma}_\mu C^\mu + \sigma^\mu \Cb_\mu \right\} - W_k[J;\gb] - \Delta S_k[\Phi;\gb] \, .
\ee
Here we used $\Phi = (h,\Cb_\mu,C^\mu)$ to denote the collection of expectation values.

The key property of the effective average action \eqref{def.eaagrav} is that its dependence on the coarse-graining scale $k$ is again governed by a formally exact functional renormalization equation taking the form \eqref{FRGE}. Its derivation essentially follows the one for the scalar theory. Taking the derivative of \eqref{def.eaagrav} with respect to the RG time $t$ and expressing the right-hand side in terms of the Hessian of $\Gamma_k[\Phi;\gb]$ one finds \cite{Reuter:1996cp}
\be\label{frge.grav1}
\begin{split}
\p_t \Gamma_k[\Phi;\gb] = & \, \frac{1}{2} {\rm Tr}\left[ \left( \Gamma_k^{(2)} + \cR_k \right)^{-1}_{hh} \left( \p_t \cR_k \right)_{hh} \right] \\
 & - \frac{1}{2} {\rm Tr} \left[ \left\{ \left(\Gamma_k^{(2)} + \cR_k \right)^{-1}_{\Cb C} - \left( \Gamma_k^{(2)} + \cR_k \right)^{-1}_{C\Cb} \right\} (\p_t \cR_k)_{\Cb C}\right] \, . 
\end{split}
\ee
Here the matrix elements constituting the Hessian of $\Gamma_k$ are defined via
\be\label{eq.hessians}
\left( \Gamma_k^{(2)} \right)_{ij}(x,y) \equiv \frac{1}{\sqrt{\gb(x)} \sqrt{\gb(y)}} \,
\frac{\delta^2 \Gamma_k}{\delta \Phi^i(x) \delta \Phi^j(y)} \, . 
\ee
For the Grassmann-valued (anti-commuting) fields in the ghost sector, we adopt the convention that matrix elements are defined in terms of left-derivatives, i.e.,
\be\label{eq.hessianghost}
\left( \left( \Gamma_k^{(2)} \right)_{\Cb C} \right)_\mu{}^\nu(x,y) = \frac{1}{\sqrt{\gb(x)}} \frac{\delta}{\delta C^\mu(x)} \frac{1}{\sqrt{\gb(y)}} \frac{\delta}{\delta \Cb_\nu(y)} \, \Gamma_k[\Phi;\gb] \, . 
\ee
Introducing a supertrace STr which includes a sum over all fluctuation fields as well as a minus sign for Grassmann-valued degrees of freedom, Eq.\ \eqref{frge.grav1} can again be written in compact form,
\be\label{frge.grav2}
\p_t \Gamma_k[\Phi;\gb] =  \frac{1}{2} {\rm STr}\left[ \left( \Gamma_k^{(2)} + \cR_k \right)^{-1} \, \p_t \cR_k  \right] \, . 
\ee

This equation maintains all the properties discussed in the context of the scalar theory. It is the central result of this section and constitutes the starting point for investigating the Wilsonian renormalization group flow of gravity. Notably, its use is not limited to the case where the gravitational degrees of freedom are encoded in metric fluctuations. It is also applicable to formulations building on different sets of degrees of freedom, including unimodular gravity, the Hilbert-Palatini formulation, the Arnowitt-Deser-Misner (ADM) decomposition of the metric degrees of freedom, and also Ho\v{r}ava-Lifshitz gravity. This makes \eqref{frge.grav2} a powerful and rather universal tool to study the quantum properties of gravity beyond perturbation theory and its use in practical computations will be discussed in Sec.\ \ref{sec.eh}.
 
 At this point the following conceptual clarifications are in order. At first sight the introduction of a background metric seems to contradict the idea of background independence intrinsic to general relativity. This is not the case though. Keeping $\gb_{\mu\nu}$ generic essentially corresponds to quantizing the theory in all backgrounds simultaneously. Subsequently, one can then evoke a dynamical principle determining $\gb_{\mu\nu}$. In this way one retains background independence even in the presence of a background metric. This viewpoint underlies the concept of self-consistent backgrounds developed in \cite{Becker:2014pea,Pagani:2019vfm}.

\subsection{Common approximation schemes}
\label{sect.34}
The Wetterich equation \eqref{frge.grav2} constitutes a formally exact equation. Finding exact solutions to it is equivalent to carrying out the functional integral \eqref{eq.Zdef}. This is extremely ambitious though and usually can not be carried out exactly. Thus, one has to resort to approximations.

Probably, the most prominent approximation is perturbation theory. In this case the standard result is recovered by neglecting the $k$-dependence of $\Gamma_k^{(2)}$ on the right-hand side of Eq.\ \eqref{frge.grav2} and approximating $\Gamma_k^{(2)} \rightarrow S_\Lambda^{(2)}$ with $S_\Lambda$ the bare action defined at the UV-scale $\Lambda$. This approximation turns the trace into a total derivative
\be
\p_t \Gamma_k \simeq \frac{1}{2} \p_t {\rm Tr} \left[ \ln \left( S_\Lambda^{(2)} + \cR_k \right) \right] \, . 
\ee
Here and in the following we use $\simeq$ to indicate an approximation of the exact flow.
Integrating this equation from the UV-scale down to $k=0$ and assuming that the regulator vanishes at the boundaries then yields the standard formula for the one-loop effective action
\be
\Gamma^{\rm 1-loop} = S_\Lambda + \frac{1}{2} {\rm Tr} \left[ \ln S_\Lambda^{(2)} \right] \, . 
\ee

The investigation of RG fixed points typically builds on non-perturbative approximation schemes though. The basic idea is to start from the exact flow and project it onto a subspace spanned by a finite (or even infinite) set of interaction monomials $\cO_i$. In the setup introduced in Sec.\ \ref{sec.ea}, this amounts to truncating the sum in eq.\ \eqref{eq.expansionG} to a finite set
\be
\Gamma_k \simeq \sum_{i=1}^N \, \bar{u}_i(k) \, \cO_i \, .
\ee
These types of approximations can be set up systematically, either in the form of a derivative expansion or a vertex expansion. These commonly used non-perturbative approximation schemes will be discussed in Sects.\ \ref{sect.341} and \ref{sect.342}, respectively.
\subsubsection{Derivative and curvature expansion}
\label{sect.341}
When developing non-perturbative approximation schemes, it is important to appreciate that $\Gamma_k$ depends on two metric arguments $g_{\mu\nu}$ and $\gb_{\mu\nu}$. The dependence on $g_{\mu\nu}$ can be traded for the fluctuations $h_{\mu\nu}$ by substituting the linear split \eqref{eq.backgrounddecomp}. Structurally, it is then convenient to organize the contributions in $\Gamma_k$ according to their transformation properties with respect to the background and quantum gauge transformations
\be\label{eq.gammaexp}
\Gamma_k[g, \gb, \bar{C}, C] = \bar{\Gamma}_k[g] + \widehat{\Gamma}_k[g,\gb] + \Gamma^{\rm gf}_k[g,\gb] + \Gamma_k^{\rm ghost}[g, \gb, \bar{C}, C] \, . 
\ee 
Here $\Gamma_k^{\rm gf}[g,\gb]$ and $\Gamma_k^{\rm ghost}[g, \gb, C, \bar{C}]$ are the standard gauge-fixing and ghost terms. The subscript $k$ thereby indicates that these sectors can contain $k$-dependent couplings, as, e.g., a wave-function renormalization for the ghost fields. The contribution $\bar{\Gamma}_k[g]$ collects all terms constructed from $g_{\mu\nu}$ only. By construction, $\bar{\Gamma}_k[g]$ is then invariant with respect to both background and quantum gauge transformations
\be
\delta^B \bar{\Gamma}_k[g] = 0 \, , \qquad \delta^Q \bar{\Gamma}_k[g] = 0 \, . 
\ee
The terms contained in $ \widehat{\Gamma}_k[g,\gb]$ genuinely depend on both arguments. It collects the ``off-diagonal'' contributions and satisfies
\be
 \widehat{\Gamma}_k[g,g] = 0 \, .
\ee

A rather broad class of approximations based on \eqref{eq.gammaexp} truncates the effective average action by setting $ \widehat{\Gamma}_k[g,\gb] \simeq 0$. Commonly, these approximations are referred to as \emph{single-metric approximations} \cite{Manrique:2009uh,Manrique:2010am,Manrique:2010mq}. Most approximations along these lines also work with a classical ghost sector, setting $\Gamma_k^{\rm ghost}[g, \gb, \bar{C}, C] \simeq S^{\rm ghost}[g, \gb, \bar{C}, C]$. 

Building on the results by Fulling, King, Wybourne, Cummins \cite{Fulling:1992vm} (further elaborated on in \cite{Decanini:2008pr}), one can systematically construct a basis $\cO_i[g]$ in which $\bar{\Gamma}_k[g]$ can be expanded. The explicit construction of the independent basis elements needs to take into account redundancies due to the Bianchi identity $D_{[\mu} R_{\alpha\beta]\gamma\delta} = 0$. In addition, low-dimensional cases are subject to additional simplifications, e.g., due to the vanishing of the Weyl tensor in $d=3$.

 The symmetries of $\bar{\Gamma}_k[g]$ dictate that the corresponding monomials are built from the Riemann tensor $R_{\mu\nu\rho\sigma}$, its contractions, and covariant derivatives $D_\mu$ acting on the curvature tensors. Convenient building blocks for the basis elements are then provided either by the Riemann basis
\be\label{eq.riemannbasis}
\cO_i[g] = \cO_i[\sqrt{g},R,R_{\mu\nu},R_{\mu\nu\rho\sigma},D_\mu]
\ee
or the Weyl basis
\be\label{eq.weylbasis}
\cO_i[g] = \cO_i[\sqrt{g},R,R_{\mu\nu},C_{\mu\nu\rho\sigma},D_\mu]
\ee
The two choices are related by the identity
\be\label{RiemannToWeyl}
C_{\mu\nu\rho\sigma} = R_{\mu\nu\rho\sigma} - \frac{2}{d-2}\left( g_{\mu[\rho} R_{\sigma]\nu} - g_{\nu[\rho} R_{\sigma]\mu} \right) + \frac{2}{(d-1)(d-2)} R g_{\mu[\rho} g_{\sigma]\nu} \, . 
\ee
In terms of structural aspects, it is often useful to work in the Weyl basis, since this choice disentangles the contributions of the higher-derivative terms to the flat-space graviton propagator.

The expansion of $\bar{\Gamma}_k[g]$ can be organized systematically by counting the number of spacetime derivatives $n$ contained in the monomial $\cO_i[g] \equiv \int d^dx \sqrt{g} \tilde{\cO}_l^n[g]$. The index set $\{i\} \mapsto \{n,l\}$ where $l$ enumerates the basis elements occurring at a fixed order $n$, see Table \ref{tab.derivativeexp} for examples. This scheme is called the \emph{derivative expansion} of $\bar{\Gamma}_k[g]$. The basis elements appearing at the lowest orders are given in Table \ref{tab.derivativeexp}.
\begin{table}[t!]\centering
	\renewcommand{\arraystretch}{1.5}
			\begin{tabular}{c|cc|ccc}
				\backslashbox{$n$}{$l$}
			  & \hspace{7mm} $1$ \hspace{7mm} & \hspace{7mm} $2$ \hspace{7mm} & \hspace{7mm} $3$ \hspace{7mm} & \hspace{7mm} $4$ \hspace{7mm} & \hspace{7mm} $\cdots$ \hspace{7mm} \\ \hline \hline
			$\vdots$ & 	$\vdots$ & 	$\vdots$ & 	$\vdots$ & 	$\vdots$ & 	$\vdots$ \\
		\hspace{3mm} $8$ \hspace{3mm} & $R \Delta^2 R$ & $C_{\mu\nu\rho\sigma} \Delta^2 C^{\mu\nu\rho\sigma}$ & $R^4$ & $R^2 \, R_{\mu\nu} \, R^{\mu\nu}$ & $\cdots$ \\
			$6$ & $R \Delta R$ & $C_{\mu\nu\rho\sigma} \Delta C^{\mu\nu\rho\sigma}$ & $R^3$ & $R \, R_{\mu\nu} \, R^{\mu\nu}$ & $+6$ more  \\
			$4$ & $R^2$ & $C_{\mu\nu\rho\sigma} C^{\mu\nu\rho\sigma}$ & $E$ & & \\
			$2$ & $R$ & & & & \\
			$0$ & $\mathbb{1}$ & & & & \\ \hline \hline
			\end{tabular} 
	\caption{Illustration of the interaction monomials $\tilde{\cO}_l^n[g]$ appearing in the derivative expansion of $\bar{\Gamma}_k[g]$ at order $n$ using the Weyl basis \eqref{eq.weylbasis}. The terms listed in the middle contribute to the graviton propagator in a four-dimensional flat background. Terms in the right-most block contribute terms proportional to the background curvature in  $\Gamma_k^{(2)}[h=0;\gb]$ and may be interpreted as ``potential terms''. Furthermore, $E= R_{\mu\nu\rho\sigma} R^{\mu\nu\rho\sigma} - 4 R_{\mu\nu}R^{\mu\nu} + R^2$ denotes the integrand of the Gauss-Bonnet term, which is topological in $d=4$.}\label{tab.derivativeexp}
\end{table}
The number of independent basis elements increases significantly with each order in the derivative expansion. This expansion scheme provides a good ordering principle  when studying the ``low-energy'' properties of the theory. For fixed points which are Gaussian or ``almost-Gaussian'', the power-counting also provide a good guiding principle whether a given operator is relevant or irrelevant. 

A conceptual shortcoming of the derivative expansion is that truncating the series of terms contributing to the gravitational propagator induces potentially spurious poles  \cite{Becker:2017tcx}. The reason is that the approximation intrinsic to the derivative expansion leads to inverse propagators which are polynomial in the momentum. Hence it is difficult to address questions about stability and the potential presence of ghosts within this approximation \cite{Platania:2020knd,Platania:2022gtt}.

As stressed in \cite{Knorr:2019atm}, this feature can be bypassed by switching to a curvature expansion. The basic idea is to collect the covariant derivatives appearing in interaction monomials in operator-valued functions, called \emph{form factors}. These capture the dependence of propagators and interaction vertices on the (generalized) momenta of the fields and can also be defined in an arbitrary curved background spacetime. Building on the examples given in Table \ref{tab.derivativeexp}, the form factors appearing at the lowest non-trivial order in the curvature expansion arise from combining the terms in the columns with $l=1$ and $l=2$:
\be\label{eq.formfactors}
\begin{split}
& \sum_{i=0} \bar{u}^i(k) \, R \, \Delta^n \, R \mapsto R \, W^R_k(\Delta) \, R \, , \\
& \sum_{i=0} \bar{u}^i(k) \, C_{\mu\nu\rho\sigma} \, \Delta^n \, C^{\mu\nu\rho\sigma} \mapsto C_{\mu\nu\rho\sigma} \, W^C_k(\Delta) \, C^{\mu\nu\rho\sigma} \, . 
\end{split}
\ee
Notably, there are only two form factors appearing at second order in the spacetime curvature. A potential third function $R_{\mu\nu} \, W^{\rm Ric}_k(\Delta) \, R^{\mu\nu}$ can be mapped to \eqref{eq.formfactors} and higher-curvature terms by applying the Bianchi identity. The functions $W^C_{k=0}(\Delta)$ and $W^R_{k=0}(\Delta)$ fix the graviton propagator in a flat background. From Table \ref{tab.derivativeexp}, it is also apparent that there is no form factor at first order in the curvature expansion. Any derivatives acting on $R$ would lead to a surface term. As a consequence, Newton's constant $G_0$ (and also the cosmological constant $\Lambda_0$) can not carry a dependence on the physical momenta of the field.

The $k$-dependence of a form factor can again be obtained by substituting the corresponding ansatz for $\Gamma_k$ into the Wetterich equation and projecting the flow on the corresponding subspace. In general, this results in a non-linear integro-differential equation for the unknown functions, see Table \ref{tab.diffeqs}. Solving these equations either numerically or by employing pseudospectral methods then allows to obtain information on the graviton propagator and momentum dependence of interaction vertices, see \cite{Bosma:2019aiu} for pioneering work in this direction.
\begin{table}[t!]
\centering
\renewcommand{\arraystretch}{1.7}
\begin{tabular}{ccc}
approximation of $\Gamma_k$ & \hspace{3mm} structure of RG flow \hspace{3mm} & \hspace{3mm} fixed points \hspace{3mm} \\ \hline \hline
finite number of $\cO_i$ & ODEs & algebraic \\ 
\begin{tabular}{@{}c@{}}field dependent functions \\[-1.4ex] $f_k(R_1, \cdots, R_n)$\end{tabular} &
\begin{tabular}{@{}c@{}}PDEs \\[-1.4ex] $(n+1)$ variables\end{tabular} & 
\begin{tabular}{@{}c@{}}PDEs \\[-1.4ex] $n$ variables\end{tabular} \\
\begin{tabular}{@{}c@{}}momentum-dependent form factors \\[-1.4ex] $f_k(p_1, \cdots, p_n)$\end{tabular} &
\begin{tabular}{@{}c@{}}IDEs \\[-1.4ex] $(n+1)$ variables\end{tabular} & 
\begin{tabular}{@{}c@{}}IDEs \\[-1.4ex] $n$ variables \end{tabular} \\ \hline \hline
\end{tabular}
\caption{Summary of the mathematical structures capturing the flow of $\Gamma_k$
in different classes of approximations. Depending on the scale-dependent terms
retained in $\Gamma_k$, the projected flow equations are non-linear ordinary differential
equations (ODEs), partial differential equations (PDEs), or (partial) integro-
differential equations (IDEs). Since fixed functionals are $k$-stationary solutions, their
structure is encoded in differential equations which contain one variable less than the
\mbox{corresponding} flow equation.}\label{tab.diffeqs}
\end{table}
\subsubsection{Incorporating higher-order interaction vertices}
\label{sect.342}
The background approximation evaluates the Wetterich equation at zeroth order in the fluctuation field. This class of approximations can then be extended systematically by taking into account higher orders of the fluctuation field. This is the idea behind the bimetric computations initiated in \cite{Manrique:2009uh,Manrique:2010am,Manrique:2010mq} and the fluctuation approach reviewed in \cite{Pawlowski:2020qer}. It can be implemented systematically by performing a vertex expansion of $\Gamma_k[h;\gb]$ in powers of the fluctuation field:\footnote{The discussion of the ghost contributions follows the same lines, but is suppressed for the sake of readability.}
\be\label{eq.vertex}
\Gamma_k[h;\gb] = \sum_{n,l} \frac{1}{n!} \int d^dx \; \Gamma_k^{l;\mu_1\nu_1 \cdots \mu_n \nu_n}[\gb] \; h_{\mu_1\nu_1} \cdots h_{\mu_n\nu_n} \, . 
\ee
Here $l$ enumerates the set of independent tensor structures contracting $n$ powers of the fluctuation fields. Note that all dependence on the background metric is stored in $\Gamma_k^{l;\mu_1\nu_1 \cdots \mu_n \nu_n}[\gb]$. Similarly to \eqref{eq.riemannbasis} and \eqref{eq.weylbasis}, the vertices can be build from $\sqrt{\gb}$, background curvature tensors and their contractions, as well as the background covariant derivative. By construction $\Gamma_k^{l;\mu_1\nu_1 \cdots \mu_n \nu_n}[\gb]$ transforms as a tensor of the corresponding rank with respect to background gauge transformations. Since the expansion captures contributions from both $\bar{\Gamma}_k[g]$ and $\widehat{\Gamma}_k[g,\gb]$, quantum gauge invariance is broken and the classification of admissible vertices is significantly more complicated than in the single-metric case. Prototypical examples of terms appearing in the vertex expansion can be obtained from expanding the gauge-fixed Einstein-Hilbert action in powers of $h_{\mu\nu}$. Explicit examples can then be found in Eqs.\ \eqref{eq.gammaquad2} and \eqref{eq.flucttensors}. 

The $k$-dependence of the vertices appearing in \eqref{eq.vertex} can again be obtained from the Wetterich equation. Taking functional derivatives of \eqref{eq.Wetterich} with respect to the fluctuation fields gives a hierarchy of equations determining of the schematic form
\be\label{eq.vertexflow}
\p_t \Gamma^{(n)}_k[\gb] = \text{Flow}\left[ \Gamma^{(2)}_k[\gb], \cdots, \Gamma^{(n+2)}_k[\gb] \right] \, . 
\ee 
Here the superscript indicates the $n$-th functional derivative of $\Gamma_k$ with respect to the fluctuation fields, cf.\ \eqref{eq.hessians}. Background computations evaluate this hierarchy at zeroth order in $n$. Note that the right-hand side also depends on the higher-order vertices $\Gamma^{(n+1)}_k[\gb]$ and $\Gamma^{(n+2)}_k[\gb]$. The truncation of the system to a finite set of tensor structures then requires an assumption on these higher-order vertices in order to close the system. A typical strategy is to approximate the couplings appearing at the orders $(n+1)$ and $(n+2)$ by the ones appearing at the lower orders in the hierarchy. 

In practice, computations maintaining information about the fluctuation fields have mainly been carried out in a flat background, setting $\gb_{\mu\nu} = \delta_{\mu\nu}$. This choice gives access to powerful momentum space techniques and the hierarchy \eqref{eq.vertexflow} can then be evaluated by employing standard Feynman diagram techniques. In particular, eq.\ \eqref{eq.vertex} simplifies to
\be\label{eq.vertex2}
\Gamma_k[h;\delta] = \sum_{n,l} \frac{1}{n!} \left( \prod_n \int \frac{d^dp}{(2\pi)^d} \right) \, \Gamma_k^{l;\mu_1\nu_1 \cdots \mu_n \nu_n}(p_1,\cdots, p_n) h_{\mu_1\nu_1}(p_1) \cdots h_{\mu_n\nu_n}(p_n) \, , 
\ee  
where the $p_i$ are the momenta of the fluctuation fields. This has led to significant insights on the momentum-dependence of the graviton two-point function 
\cite{Christiansen:2014raa,Bonanno:2021squ} and resolving the momentum-dependence of three- and four-point vertices \cite{Christiansen:2015rva,Denz:2016qks}.

\subsection{Further developments}
\label{sect.33}
The discussion of the Wetterich equation and its properties mainly followed the initial constructions \cite{Wetterich:1992yh,Reuter:1993kw,Reuter:1996cp}. We complete our exposition by briefly introducing two recent developments, the \emph{minimal essential scheme} \cite{Baldazzi:2021ydj} (Sec.\ \ref{sect.331}) and the $N$-type cutoffs \cite{Becker:2020mjl,Becker:2021pwo} (Sec.\ \ref{sect.332}).
\subsubsection{The minimal essential scheme}
\label{sect.331}
Ultimately, the goal of the gravitational asymptotic safety program is the construction of observables. From this perspective, it turns out that the theory space introduced in Sec.\ \ref{sec.ea}, spanned by all possible interaction monomials $\cO_i$, contains redundancies in the sense that not all couplings appearing in this basis will also enter into the observables. A prototypical example is the wave-function renormalization of a field, which drops out from the construction of scattering amplitudes. On this basis one distinguishes between \emph{essential couplings} which enter into the expressions for physical observables and \emph{inessential couplings} whose values can be changed without affecting the predictions of the theory.

Typically, a change in an inessential coupling can be absorbed into a reparameterization of the dynamical variables. Considering an infinitesimal change in the field, $\chi \mapsto \chi + \xi[\chi]$, the underlying action transforms as\footnote{We use the ``$\cdot$'' to indicate an integral over spacetime and potentially a sum over internal indices labeling the fields.}
\be\label{fieldrep}
S[\chi] \mapsto S[\chi] + \xi[\chi] \cdot \frac{\delta}{\delta \chi} S[\chi] \, . 
\ee
This underlies the general statement that operators which are proportional to the equations of motion can be removed by a field redefinition and are thus linked with inessential couplings \cite{tHooft:1973pz}. Generically, one can also consider finite frame transformations to a new field parameterization,
\be\label{eq.frame1}
\phi(x) = \phi[\chi](x),
\ee
requiring that the map is quasi-local and invertible.

Implementing the procedure of removing inessential couplings at the level of the functional renormalization group is slightly more complicated. Since the corresponding couplings depend on the coarse-graining scale $k$, the field-redefinitions required in this process inherit this scale-dependence. Thus the frame transformation \eqref{eq.frame1} is promoted to be $k$-dependent
\be\label{eq.frame2}
\phi_k(x) = \phi_k[\chi](x),
\ee
This effect can be accommodated by formulating the Wetterich equation in a frame-covariant way \cite{Pawlowski:2005xe}
\be\label{FRGE.framecov}
\left( \p_t + \Psi_k[\phi] \, \frac{\delta}{\delta \phi} \right) \Gamma_k[\phi]
= \frac{1}{2} {\rm Tr} \left[ \left(\Gamma^{(2)}_k + \cR_k \right)^{-1} \left( \p_t + 2 \,  \Psi_k[\phi] \frac{\delta}{\delta \phi} \right) \cR_k \right] \, . 
\ee
The renormalization group kernel
\be\label{RGkernel}
\Psi_k[\phi] \equiv \p_t \phi_k[\chi]
\ee
thereby accounts for the $k$-dependence of the frame transformation.

In order to illustrate the working of the minimal essential scheme, we return to the example of a scalar field theory. Explicitly, we set
\be\label{eq.gammaex1}
\Gamma_k[\chi] = \int d^dx \left\{ \frac{Z_k}{2} \chi \left[ -\p^2 + m_k^2 \right] \chi + \frac{Z_k^2 \lambda_k}{12} \chi^4 + \cdots \right\} \, .
\ee
Here $m_k$ and $\lambda_k$ are scale-dependent couplings, $Z_k$ is the wave-function of the field, and the dots symbolizes additional interaction terms. The wave-function renormalization constitutes an inessential coupling and we seek to remove it by a $k$-dependent frame transformation.  Inspecting \eqref{eq.gammaex1} indicates that this can be achieved by a $k$-dependent frame-transformation which is linear in the field
\be
\phi_k = Z_k^{1/2} \, \chi \, . 
\ee
The kernel \eqref{RGkernel} then evaluates to $\Psi_k[\phi] = - \frac{1}{2} \eta_k \phi_k$ where $\eta_k \equiv - \p_t \ln Z_k$ is the anomalous dimension of the field. Evaluating \eqref{FRGE.framecov} then yields
\be\label{eq.gammaex2}
\left( \p_t - \frac{1}{2} \eta_k \, \phi \, \frac{\delta}{\delta \phi} \right) \Gamma_k[\phi]
= \frac{1}{2} {\rm Tr} \left[ \left(\Gamma^{(2)}_k[\phi] + \cR_k \right)^{-1} \left( \p_t \cR_k - \eta_k \, \cR_k  \right) \right] \, . 
\ee
The new functional $\Gamma_k[\phi]$ is then independent of $Z_k$. More precisely, the inessential coupling has been fixed to $Z_k=1$ at all scales. The result \eqref{eq.gammaex2} furthermore shows that $\eta_k$ depends on the essential couplings of the theory only. 

As pointed out in \cite{Baldazzi:2021ydj} and illustrated by our explicit example above, the use of the frame-covariant flow equation in combination with the minimal-essential scheme may lead to significant technical simplifications when constructing solutions to the flow equation. In practice, these simplifications can be exploited systematically by parameterizing the kernel $\Psi_k[\phi]$ in terms of $k$-dependent $\gamma$-functions \cite{Baldazzi:2021orb,Knorr:2022ilz}. The freedom gained in this way can then be used to fix the inessential coupling constants to specific values. The scale-dependence of the theory is then captured by the $\beta$-functions (governing the $k$-dependence of the essential couplings) and the $\gamma$-functions (governing the $k$-dependence of the inessential ones). Both sets of equations depend on the essential couplings only. The last property then simplifies the search for RG fixed points in a significant way.

\subsubsection{Flows in terms of $N$-type cutoffs}
\label{sect.332}
Recently, a novel regularization scheme via dimensionless $N$-type cutoffs has been introduced
\cite{Becker:2020mjl,Becker:2021pwo,Banerjee:2023ztr}, which may constitute a more physical alternative to the usually employed dimensionful UV cutoffs. The motivation for the introduction of a scale-free regularization scheme is the construction of regularized quantum systems, which have the potential of being physically realizable themselves. In this way, physical properties of the theory, which conventionally are to be studied in the quantum field theory limit, could already be probed at the level of the regularized system. Moreover, this scale-free regularization scheme is designed in a way, such that self-consistent background geometries can easily be accessed.

Schematically, the $N$-type cutoff regularizes the path integral \eqref{eq.Zdef} as follows. One expands the field in the eigenbasis of a suitable self-adjoint operator, e.g., the background Laplacian, such that the corresponding eigenvalues increase with $n \in \mathbb{N}$ (or $n \in \mathbb{R}^+$), cf.\ Eq.\ \eqref{eigenmodes}. Then the path integral is regularized by restricting the domain of integration to the field modes $h^n$ with $n \le N$. As a result, one obtains $N$-sequences of regularized quantum systems, which in principle are physically realizable.

As a first application, the self-consistent spherical background geometries stemming
from summing up vacuum energy of a scalar field \cite{Becker:2020mjl} as well as metric fluctuations \cite{Becker:2021pwo} have been studied. The striking result, which is due to background independence, is that the self-consistent scalar curvatures $R(N)$ vanished for $N \rightarrow \infty$ in both cases. This is precisely the opposite behavior of the commonly perceived cosmological constant problem, according to which the background curvature, and therewith the total cosmological constant, should diverge when removing the UV regulator. Another striking result of this regularization scheme \cite{Becker:2020mjl} is that $N$-type cutoffs give an explanation of the microscopical degrees of freedom which the Bekenstein-Hawking entropy of de Sitter space counts.

	\section{The Einstein-Hilbert truncation}
	\label{sec.eh}
We proceed by giving an explicit example, illustrating how the Wetterich equation \eqref{frge.grav2} is used to extract non-perturbative information about the gravitational RG flow. The discussion is based on the arguably simplest approximation for the effective average action $\Gamma_k$, the Einstein-Hilbert truncation. Starting from the seminal paper \cite{Reuter:1996cp}, this projection has been studied in detail in a series of works \cite{Souma:1999at,Reuter:2001ag,Lauscher:2001ya,Litim:2003vp,Gies:2015tca}. It still forms an integral part of studying the RG flow in many gravity-matter systems. The present exposition differs from the historical computations where the background metric has been set to the one of the maximally symmetric $d$-sphere $S^d$. Instead, we combine the idea of the universal RG machine \cite{Benedetti:2010nr,Groh:2011vn} with off-diagonal heat-kernel techniques \cite{Gorbar:2002pw,Gorbar:2003yt,Decanini:2005gt,Benedetti:2010nr,Codello:2012kq} and carry out the derivation of the beta functions \emph{without specifying the background metric} $\gb_{\mu\nu}$. This stresses the background-independent nature of the computation and emphasizes the modern viewpoint on evaluating the FRGE in the context of gravity. In order to keep technical complications at the minimum, we adopt the harmonic gauge. The beta functions resulting from this setting are computed in Sec.\ \ref{eh.beta} and the resulting fixed point structure and phase diagram is presented in Sec.\ \ref{eh.phase}. Results obtained by generalizing this computation by resorting to additional field decompositions and generalizing the gauge-fixing and regularization prescription have been obtained in \cite{Gies:2015tca} and corroborate the findings reviewed in this section. 
	\subsection{Deriving the beta functions}
	\label{eh.beta}
	The Einstein-Hilbert (EH) truncation works in the background approximation. Thus the flow is obtained at zeroth order in the fluctuation fields. As a consequence only terms of zeroth and second order in the fluctuations are needed in the evaluation of the Wetterich equation. The projection of the flow equation tracks the scale-dependence of the (background) Newton's coupling $G_k$ and the cosmological constant $\Lambda_k$. The gravitational part of the effective average action is approximated by the Einstein-Hilbert action
	\be\label{eq.einsteinhilbert}
	\Gamma_k^{\rm EH}[g] = \frac{1}{16\pi G_k} \int d^dx \sqrt{g} \left(-R + 2 \Lambda_k \right) \, ,
	\ee
	with the couplings depending on the coarse-graining scale $k$. In view of the upcoming computation, it is convenient to introduce the dimensionless counterparts of Newton's coupling and the cosmological constant as well as the anomalous dimension of Newton's coupling
	\be\label{dimlessvars}
	g_k \equiv k^{d-2} G_k \, , \qquad \lambda_k \equiv k^{-2} \Lambda_k \, , \qquad
	\eta_N(k) \equiv (G_k)^{-1} \p_t G_k \, .
	\ee
	Furthermore, geometrical quantities constructed from $\gb_{\mu\nu}$ are distinguished by a bar. E.g., $\Db_\mu$ is the covariant derivative constructed from the background metric.
	
	 In order to obtain well-defined propagators, $\Gamma^{\rm EH}_k$ must be supplemented by a gauge-fixing term and the corresponding ghost action. Concretely, we implement a background gauge-fixing
	\be\label{eq.gaugefix}
	\Gamma_k^{\rm gf}[h;\gb] = \frac{1}{32 \pi G_k \alpha} \int d^dx \sqrt{\gb} \, \gb^{\mu\nu} F_\mu F_\nu \, , 
	\ee
	where the gauge-fixing condition is taken to be linear in the fluctuation field
	\be\label{eq.gaugefixlinear}
	F_\mu[h;\gb] = \left[ \delta_\mu^\alpha \Db^\beta - \beta \gb^{\alpha\beta} \Db_\mu \right] \, h_{\alpha\beta} \, . 
	\ee
	Here $\alpha$ and $\beta$ are two gauge-parameters which can largely be chosen arbitrary \cite{Gies:2015tca}. The ghost action accompanying \eqref{eq.gaugefix} is found in the standard way and reads
	\be\label{eq.ghostaction}
	S^{\rm ghost}[h,\bar{C},C;\gb] = - \sqrt{2} \int d^dx \sqrt{\gb} \, \Cb_\mu \, \cM[g,\gb]^\mu{}_\nu \, C^\nu \, ,
	\ee
	with the Faddeev-Popov operator being
	\be\label{eq.FPop}
	\cM[g,\gb]^\mu{}_\nu = \gb^{\mu\rho} \Db^\sigma \left(g_{\rho\nu} D_\sigma + g_{\sigma\nu} D_\rho \right) - 2 \beta \gb^{\rho\sigma} \Db^\mu g_{\sigma\nu} D_\rho \, . 
	\ee
	Landau-type gauge fixings correspond to the limit $\alpha \rightarrow 0$ (with $\beta = 1/d$ being a preferred choice implementing the geometric gauge). The harmonic gauge adopted in the present computation sets $\alpha = 1$ (Feynman-type gauge) and $\beta = 1/2$. This has the technical advantage that all derivatives appear in the form of the background Laplace operator $\Delta = - \gb^{\mu\nu} \Db_\mu \Db_\nu$.
	
	For the background computation ahead, it suffices to know the ghost-action to second order in the fluctuation fields. Adopting harmonic gauge and evaluating $\cM[g,\gb]^\mu{}_\nu|_{g = \gb}$ shows that the relevant contributions are captured by
	\be\label{eq.ghost2}
	S^{\rm ghost}[h=0,\bar{C},C;\gb] =  \sqrt{2} \int d^dx \sqrt{\gb} \, \Cb_\mu \, \left[ \, \delta^\mu_\nu \Delta - \Rb^\mu{}_\nu \, \right] \, C^\nu \, .
	\ee
	Here, we used the commutator of two background-covariant derivatives evaluated on vectors in order to combine the last two terms in \eqref{eq.FPop} into the background Ricci scalar $\Rb_{\mu\nu}$. The approximation for the effective average action then combines the $\bar{\Gamma}_k[g]$ given in \eqref{eq.einsteinhilbert} with the gauge-fixing term \eqref{eq.gaugefix} and the ghost action \eqref{eq.ghostaction}
	\be\label{eq.ans2}
	\Gamma_k[h,\bar{C},C;\gb] \simeq \Gamma_k^{\rm EH}[g] + 	\Gamma_k^{\rm gf}[h;\gb]  + S^{\rm ghost}[h,C,\bar{C};\gb] \, . 
	\ee
	
	At this stage a comment on the projection prescription is in order. Substituting \eqref{eq.ans2} into its left-hand side and setting $g=\gb$ afterwards indicates that the scale-dependence of $G_k$ and $\Lambda_k$ can be read off from the coefficients multiplying
	\be\label{eq.projectionspace}
	\cO_0 = \int d^dx \sqrt{\gb} \, , \qquad \cO_1 = \int d^dx \sqrt{\gb} \Rb \, . 
	\ee
	All other interaction monomials spanning the gravitational theory space do not contribute to the computation. This entails the following, profound consequence. Eq.\ \eqref{eq.projectionspace} corresponds to a derivative expansion truncated at first order in the spacetime curvature. Hence all terms containing two or more curvature tensors are outside the subspace spanned by our approximation. Moreover, \eqref{eq.projectionspace} does not contain derivatives of a curvature tensor. Hence, there is no need to track such terms in the present computation. These considerations allow to formulate projection rules, stating that
	\be\label{eq.projectrule}
	\Db_\mu \Rb_{\alpha\beta\gamma\delta} \simeq 0 \, , \qquad O(\Rb^2) \simeq 0 \, . 
	\ee
	We stress that these rules should not be read as restrictions on $\gb_{\mu\nu}$. They merely identify structures which do not contribute to the computation. As a corollary of these relations, we conclude that we can freely commute covariant derivatives and curvature tensors, since the commutators just produce terms outside of the projection spanned by \eqref{eq.projectionspace}. 
	
	The first step in evaluating the trace appearing within the FRGE \eqref{frge.grav2} consists in expanding \eqref{eq.ans2} to second order in the fluctuation fields. In the ghost-sector the result is already given in \eqref{eq.ghost2}. For the gravitational fluctuations, we expand
	\be
	\Gamma_k[\gb+h,\gb] = \Gamma_k[\gb,\gb] + O(h) + \Gamma^{\rm quad}_k[h;\gb] + O(h^3) \, . 
	\ee
	 The relevant coefficient $\Gamma^{\rm quad}_k[h;\gb]$ is readily found using computer algebra packages like xAct \cite{Brizuela:2008ra} and has the form
	 \be\label{eq.gammaquad2}
	 \Gamma^{\rm quad}_k = \frac{1}{32 \pi G_k} \int d^dx \sqrt{\gb} \; \frac{1}{2} \, h_{\mu\nu} \left[ K^{\mu\nu}{}_{\alpha\beta} \left( \Delta - 2 \Lambda_k \right) + V^{\mu\nu}{}_{\alpha\beta} \right] \, h^{\alpha\beta} \, . 
	 \ee
	 Here the ``kinetic'' and ``potential'' parts have the explicit form
	 \be\label{eq.flucttensors}
	 \begin{split}
	 K^{\mu\nu}{}_{\alpha\beta} = & \, \frac{1}{2} \left( \delta^\mu_\alpha \delta^\nu_\beta + \delta^\mu_\beta \delta^\nu_\alpha - \gb^{\mu\nu} \gb_{\alpha\beta} \right) \, ,  \\
	 V^{\mu\nu}{}_{\alpha\beta} = & \, \Rb \, K^{\mu\nu}{}_{\alpha\beta}  +  \left( \gb^{\mu\nu} \Rb_{\alpha\beta} + \Rb^{\mu\nu} \gb_{\alpha\beta} \right) - 2 \delta^{(\mu}_{(\alpha} \Rb^{\nu)}_{\beta)} - 2 \Rb^{(\mu}{}_{(\alpha}{}^{\nu)}{}_{\beta)} \, . \\
	 \end{split}
     \ee
     The potential $V$ collects all terms containing the spacetime curvature and is of first order in a curvature expansion.
     
     In the next step, we would like to diagonalize the kinetic terms in the quadratic form \eqref{eq.gammaquad2}. This can be achieved by decomposing $h_{\mu\nu}$ into component fields, resorting to the transverse-traceless decomposition \cite{York:1973ia,Lauscher:2001ya}. In the present case, it suffices to split the fluctuations into their trace- and traceless part
     \be\label{eq.fielddec}
     h_{\mu\nu} = \hh_{\mu\nu} + \frac{1}{d} \gb_{\mu\nu} h \, , \qquad \gb^{\mu\nu} \hh_{\mu\nu} = 0 \, . 
     \ee
Substituting this decomposition into \eqref{eq.gammaquad2} then yields 
	\be\label{eq.gammaquad}
	\begin{split}
	\Gamma^{\rm quad}_k[h;\gb] = \frac{1}{32 \pi G_k} &  \int d^dx \sqrt{\gb} \bigg[
	\frac{1}{2} \hh_{\mu\nu} \left[ \Delta - 2 \Lambda_k + \Rb \right] \hh^{\mu\nu} \\
	& - \left(\frac{d-2}{4d} \right) \, h \, \left[ \Delta - 2 \Lambda_k + \frac{d-4}{d} \Rb \right] \, h \\
	& - \Rb_{\mu\nu} \hh^{\mu\alpha} \hh_{\alpha}{}^\nu - \Rb_{\mu\nu\alpha\beta} \hh^{\mu\alpha} \hh^{\nu\beta} + \frac{d-4}{d} \Rb_{\mu\nu} \, h \, \hh^{\mu\nu}
	\bigg] \, . 
	\end{split}
	\ee
	
	At this point, we are ready to specify the explicit form of the regulator $\cR_k$. We dress up the Laplacians according to
	\be
	\Delta \mapsto \Delta + R_k \, ,
	\ee
	where $R_k(\Delta) = k^2 R^{(0)}(\Delta/k^2)$ is the dimensionful cutoff function and $R^{(0)}(z)$ the corresponding profile. In the nomenclature of the review \cite{Codello:2008vh} this corresponds to a cutoff of type I. This choice implements the initial idea of supplying the fluctuation field with a $k$-dependent mass term. The resulting $\cR_k$ is then diagonal in field space with its matrix elements given by
	\be\label{eq.regeh}
	\cR_k^{\hh\hh} = \frac{1}{32 \pi G_k} R_k \, \unit_{2T} \, , \quad 
	\cR_k^{hh} = - \frac{1}{32 \pi G_k} \left( \frac{d-2}{2d} \right) R_k \, , \quad 
	\cR_k^{\bar{C}C} = \sqrt{2} R_k \, \unit_1 \, . 
	\ee
	Here
	\be\label{eq.units}
	\unit_{2T}^{\mu\nu}{}_{\alpha\beta} = \frac{1}{2}\left(\delta^\mu_\alpha \delta^\nu_\beta + \delta^\mu_\beta \delta^\nu_\alpha
	  \right) - \frac{1}{d} \gb^{\mu\nu} \gb_{\alpha\beta} \, , \qquad \unit_1^\mu{}_\nu = \delta^\mu_\nu \, , 
	\ee
	are the units on the space of symmetric traceless two-tensors (2T) and vectors (1), respectively.
	
	We now proceed by constructing the inverse of the regularized Hessian. For the gravitational degrees of freedom, we encounter the two-by-two matrix
	\be\label{eq.gamma2reg}
	\left[\Gamma^{(2)}_k + \cR_k \right]^{ij} =  \left[
	\begin{array}{cc}
		K_{2T}(\Delta) \, \unit_{2T} + V_{2T} & V_\times \\
		V_\times^\dagger & K_{0}(\Delta) \, \unit_{0} + V_{0}
	\end{array}
	\right] \, . 
	\ee
	Here $i,j = \{\hh,h\}$ labels the fields and we suppress all spacetime indices for the sake of readability. The explicit form of the kinetic functions $K$ and the potentials $V$ can be read off from Eqs.\ \eqref{eq.gammaquad} and \eqref{eq.ghost2} and read
	\be\label{eq.kinetic}
	\begin{split}
	K_{2T}(\Delta) = & \, \frac{1}{32 \pi G_k} \left(\Delta + R_k - 2 \Lambda_k\right) \, , \\  K_{0}(\Delta) = & \, -\frac{1}{32 \pi G_k} \left(\frac{d-2}{2d} \right) \, \left(\Delta + R_k - 2 \Lambda_k \right) \, , \\
	K_{1}(\Delta) = & \,\sqrt{2} \, \Delta  \, ,
	\end{split}
	\ee
	and
	\be\label{eq.potentials}
	\begin{split}
&	V_{2T}^{\mu\nu}{}_{\alpha\beta} =  \frac{1}{32 \pi G_k} \left(\Rb \,  \unit_{2T}^{\mu\nu}{}_{\alpha\beta} - 2 \Rb^{(\mu}_{(\alpha} \delta^{\nu)}_{\beta)} - 2 \Rb_{(\alpha}{}^{(\mu}{}_{\beta)}{}^{\nu)} \right) \, ,  \\
 &	V_0 =  - \frac{1}{32 \pi G_k} \, \left(\frac{d-2}{2d} \right) \left(\frac{d-4}{d}\right) \, \Rb \, , \\
& 	 V_\times {}_{\mu\nu} =  
\frac{1}{32 \pi G_k} \left(  \frac{d-4}{d} \right) \, \left( \Rb_{\mu\nu} - \frac{1}{d} \gb_{\mu\nu} \Rb \right) \, , \\
&  V_1{}^\mu{}_\nu = - \sqrt{2} \, \Rb^\mu{}_\nu \, .  
	\end{split}
	\ee
	
	Constructing the inverse of \eqref{eq.gamma2reg} builds on the exact inversion formula for block matrices
	\be\label{eq.matinv}
	\left[ 
	\begin{array}{cc}
		A & B \\
		C & D
	\end{array}
	\right]^{-1} =
	\left[
	\begin{array}{cc}
		\left(A - B D^{-1} C \right)^{-1} & - A^{-1} B \left( D-CA^{-1} B \right)^{-1} \\
		-D^{-1} C \left(A-B D^{-1} C \right)^{-1} & \left(D - C A^{-1} B \right)^{-1}
	\end{array}
	\right] \, .
	\ee
	Since the potentials \eqref{eq.potentials} contain at least one power of the spacetime curvature, each entry can be constructed as a power series in $V$. The projection prescription \eqref{eq.projectrule} then indicates that it is sufficient to retain the terms up to one power of $V$. This implies, in particular, that the off-diagonal terms $V_\times$ do not enter into the present computation since they start to contribute at second order in $V$ only. Taking into account that the regulator $\p_t \cR_k$ is diagonal in field space, it is sufficient to consider the diagonal entries in \eqref{eq.matinv}. Explicitly, the corresponding inverses are given by
	\be\label{eq.propagators}
	\begin{split}
(32 \pi G_k)^{-1}	\left[ \Gamma_k^{(2)} + \cR_k \right]^{-1}_{\hh\hh} \simeq & \, \frac{1}{K_{2T}} - \frac{1}{K_{2T}} V_{2T} \frac{1}{K_{2T}} + O(V^2)
\, , \\
(32 \pi G_k)^{-1}	\left[ \Gamma_k^{(2)} + \cR_k \right]^{-1}_{hh} \simeq & \, \frac{1}{K_{0}} - \frac{1}{K_{0}} V_{0} \frac{1}{K_{0}} + O(V^2) \, .
	\end{split}
	\ee
	 Based on these preliminary considerations, we can now write down the projected flow equation
	\be\label{eq.traces}
	\begin{split}
	\p_t \Gamma_k = & \, \frac{1}{2} {\rm Tr}_{2T} \left[ \frac{1}{K_{2T}} \p_t \cR_k^{\hh\hh} \right] - \frac{1}{2} {\rm Tr}_{2T} \left[\frac{1}{K_{2T}} V_{2T}  \frac{1}{K_{2T}}  \p_t \cR_k^{\hh\hh} \right] \\
	& \, + \frac{1}{2} {\rm Tr}_{0} \left[ \frac{1}{K_{0}}  \p_t \cR_k^{hh} \right] - \frac{1}{2} {\rm Tr}_{0} \left[ \frac{1}{K_{0}} V_{0}  \frac{1}{K_{0}} \p_t \cR_k^{hh}  \right] \\
& 	\, - {\rm Tr}_{1} \left[ \frac{1}{K_1}  \p_t \cR_k^{\bar{C}C} \right] + {\rm Tr}_{1} \left[  \frac{1}{K_1} V_1  \frac{1}{K_1}  \p_t \cR_k^{\bar{C}C} \right] \, . 
	\end{split}
	\ee
	Here the subscripts $s = \{ 2T, 0, 1\}$ indicate that the traces are over traceless, symmetric matrices, scalars, and vectors, respectively.
	
	Structurally, the traces \eqref{eq.traces} can be separated in traces without and with operator insertion $V$. In order to evaluate the resulting expressions of the first type, we use the early-time expansion of the heat-kernel \cite{Vassilevich:2003xt}
	\be\label{eq.heatearly}
	{\rm Tr}_s\left[ e^{-s \Delta} \right] = \frac{1}{(4\pi s)^{d/2}} \, {\rm tr}(\unit_s) \, \int d^dx \sqrt{\gb} \left( 1 + \frac{1}{6} s \Rb \right) + O(\Rb^2) \, .  
	\ee
	The trace tr$(\unit_s)$ counts the number of independent field components in each sector, i.e.,
	\be
	\begin{split}
	{\rm tr}(\unit_0) = 1 \, , \quad	
	{\rm tr}(\unit_1) = d \, , \quad	
	{\rm tr}(\unit_{2T}) = \frac{1}{2}(d-1)(d+2) \, .
	\end{split}
	\ee
	The heat-kernel \eqref{eq.heatearly} can readily be extended to traces including functions of the Laplacian $W(\Delta)$. Formally introducing the (inverse) Laplace transform $\widetilde{W}(s)$ through $W(z) = \int_0^\infty ds \, \widetilde{W}(s) \, e^{-sz}$, we write
	\be\label{eq.trw}
	{\rm Tr}_s\left[W(\Delta)\right] = \int_0^\infty ds \, \widetilde{W}(s) \, {\rm Tr}_s\left[e^{-s\Delta}\right] \, . 
	\ee
	Substituting the early-time expansion \eqref{eq.heatearly}, then yields
	\be\label{tr.Ws}
	{\rm Tr}_s\left[W(\Delta)\right] = \frac{1}{(4\pi)^{d/2}} \, {\rm tr}(\unit_s) \, \int d^dx \sqrt{\gb} \, \left( Q_{d/2}[W] \,    + \frac{1}{6} \, Q_{d/2-1}[W] \,  \Rb \right) + O(\Rb^2) \, , 
	\ee
	where the $Q$-functionals are defined by
	\be
	\begin{split}
		Q_n[W] \equiv \int_0^\infty ds \, s^{-n} \, \widetilde{W}(s) \, . 
	\end{split}
    \ee
    These functionals can be re-written in terms of the original function $W(z)$:
    \be
    \begin{split}
    	Q_n[W] = & \, \frac{1}{\Gamma(n)} \int_0^\infty dz \, z^{n-1} W(z) \, , \qquad n > 0 \, , \\
    	Q_0[W] = & W(0) \, .
    \end{split}
   \ee
   For $n < 0$ one can always choose an integer $k$ such that $n+k > 0$. Integrating by parts, one then establishes that 
   \be
   Q_n[W] = \frac{(-1)^k}{\Gamma(n+k)} \int_0^\infty dz \, z^{n+k-1} \, W^{(k)}(z) \, , \qquad n < 0, \quad n+k > 0 \, . 
   \ee
   
 At this point we note that the functions $W(z)$ appearing in \eqref{eq.traces} have the generic form
 \be\label{eq.Wgeneric}
 W(z) =  \frac{G_k}{(z +R_k + w)^p} \, \p_t \left(\frac{1}{G_k} R_k \right) \, .
 \ee
  In this case, it is then convenient to trade the dimensionful $Q$-functionals with the dimensionless threshold functions
    \be\label{def.threshold}
    \begin{split}
    \Phi^p_n(w) \equiv & \, \frac{1}{\Gamma(n)} \int_0^\infty dz \, z^{n-1} \frac{R^{(0)} - z R^{(0)^\prime}(z)}{(z + R^{(0)} + w)^p} \, , \\
\widetilde{\Phi}^p_n(w) \equiv & \, \frac{1}{\Gamma(n)} \int_0^\infty dz \, z^{n-1} \frac{R^{(0)}(z)}{(z + R^{(0)}(z) + w)^p} \, ,    
\end{split}
    \ee
    where $R^{(0)}(p^2/k^2)$ is the dimensionless profile function associated with the regulator $R_k(p^2) = k^2 R^{(0)}(p^2/k^2)$. It is then readily verified that
    \be\label{Q-master}
    Q_n\left[ \frac{G_k}{(z +R_k + w)^p} \, \p_t \left(\frac{1}{G_k} R_k \right) \right] = k^{2(n-p+1)} \left( 2 \Phi^p_n(w/k^2) - \eta_N \widetilde{\Phi}^p_n(w/k^2) \right) \, , 
    \ee
    where the anomalous dimension $\eta_N$ has been introduced in \eqref{dimlessvars}. For the traces in the ghost sector, the $G_k$-dependence in \eqref{eq.Wgeneric} is absent, so that the terms proportional to $\eta_N$ do not appear in this sector.

    Starting from \eqref{eq.traces} the traces without potential insertions are readily evaluated by combining Eq.\ \eqref{tr.Ws} with the result for the Q-functionals \eqref{Q-master}. The traces including the insertion of a potential can be evaluated along the same lines. Formally, such traces can be evaluated using the off-diagonal heat-kernel formulas provided in \cite{Benedetti:2010nr}. Since the potentials \eqref{eq.potentials} do not contain any covariant derivatives and, owed to the projection prescription \eqref{eq.projectrule}, can be treated as covariantly constant leads to significant simplifications though. In this case the relevant contributions are given by the leading term in the early-time expansion \eqref{eq.heatearly} with ${\rm tr}\left[\unit_s\right] \rightarrow {\rm tr}\left[V_s\right]$. A brief computation establishes that 
   \be\label{trint}
   \begin{split}
   	{\rm tr}_{1}\left[V_{1}\right] = & - \sqrt{2} \Rb \, , \\
   	{\rm tr}_{2T}\left[V_{2T}\right] = & \frac{1}{32 \pi G_k} \frac{(d+2)(d^2-3d+4)}{2d} \, \Rb \, . 
     \end{split}
   \ee
   The remaining terms in \eqref{eq.traces} are then found by pulling the contributions \eqref{trint} out of the operator trace and evaluating the latter by again combining Eqs.\ \eqref{tr.Ws} and \eqref{Q-master}. In this way one obtains the explicit form of the right-hand side of \eqref{eq.traces}. Reading off the coefficients multiplying the interaction monomials \eqref{eq.projectionspace} gives the equations governing the scale-dependence of the dimensionful couplings $G_k$ and $\Lambda_k$.  
   
   In order to study the renormalization group fixed points of the system, it is then natural to convert the dimensionful couplings to their dimensionless counterparts \eqref{dimlessvars}. The scale-dependence of $g_k$ and $\lambda_k$ is encoded in the beta functions 
   \be\label{eq.betadef}
   \p_t g_k  = \beta_g(g_k,\lambda_k) \, , \qquad \p_t \lambda_k = \beta_\lambda(g_k,\lambda_k) \, . 
   \ee
   The explicit computation yields \cite{Reuter:1996cp}
   \be\label{eq.betafinal}
   \begin{split}
   	\beta_g(g,\lambda) = & \left(d-2+ \eta_N \right) \, g \, , \\
   	\beta_\lambda(g,\lambda) = & - (2-\eta_N) \lambda  + \frac{g}{2(4\pi)^{d/2-1}}\Big( 
   	2d(d+1) \Phi^1_{d/2}(-2\lambda) \\ & \qquad \qquad - 8d \Phi^1_{d/2}(0) - d(d+1) \eta_N \widetilde{\Phi}^1_{d/2}(-2\lambda) 
   	\Big) \, .
   \end{split}
   \ee
   The anomalous dimension $\eta_N$ takes the form
   \be
   \eta_N(g,\lambda) = \frac{g B_1(\lambda)}{1-g B_2(\lambda)} \, , 
   \ee
   with
   \be
   \begin{split}
   	B_1(\lambda) = & \frac{1}{3} (4\pi)^{1-d/2} \Big(
   	d(d+1) \Phi^1_{d/2-1}(-2\lambda) -6d(d-1) \Phi^2_{d/2}(-2\lambda) 
   	\\ & \qquad \qquad \qquad 
   	- 4d \Phi^1_{d/2-1}(0) -24 \Phi^2_{d/2}(0) \Big) \, , \\
   	B_2(\lambda) = & - \frac{1}{6} (4\pi)^{1-d/2} \Big(
   	d(d+1) \widetilde{\Phi}^1_{d/2-1}(-2\lambda) - 6d(d-1) \widetilde{\Phi}^2_{d/2}(-2\lambda)
   	\Big) \, . 
   \end{split}
   \ee
   
   At this point we have completed the explicit derivation of the beta functions \eqref{eq.betafinal} governing the scale-dependence of $\{g_k,\lambda_k\}$. The result agrees with the initial derivation \cite{Reuter:1996cp}, employing a maximally symmetric background. The present derivation shows, however, that \emph{this result is actually background independent}. Apart from general properties related to the existence of the heat-kernel \eqref{eq.heatearly}, we never specified an explicit background $\gb_{\mu\nu}$. Assuming its mere existence is sufficient to arrive at the final result.
   
	\subsection{Fixed points, RG trajectories, and phase diagram}
	\label{eh.phase}
	The beta functions \eqref{eq.betadef} encode the dependence of Newton's coupling and the cosmological constant on the coarse-graining scale $k$. These have been derived for a generic regulator $R_k$. In order to investigate the resulting fixed point structure and phase diagram, we specify the regulator to be of Litim-type \eqref{Rlitim}. In this case, the integrals appearing in the threshold functions \eqref{def.threshold} can be evaluated analytically, yielding
	\be\label{eq.threshlitim}
	\Phi^p_n(w)^{\rm Litim} = \frac{1}{\Gamma(n+1)} \frac{1}{(1+w)^p} \, , \quad 
	\widetilde{\Phi}^p_n(w)^{\rm Litim} = \frac{1}{\Gamma(n+2)} \frac{1}{(1+w)^p} \, . 
	\ee
	Upon substituting these expressions, the flow of $\{g_k,\lambda_k\}$ is governed by the  coupled, non-linear, autonomous, first-order differential equations \eqref{eq.betadef} with
	\be\label{beta-litim}
	\begin{split}
		\beta_g = & \left(2 + \eta_N \right) \, g \, , \\
	\beta_\lambda = & - (2-\eta_N) \lambda  + \frac{g}{8\pi}\Big( 
	\frac{20}{1-2\lambda}  - 16 - \frac{5}{3} \, \eta_N \,  \frac{1}{1-2\lambda} 
	\Big) \, .
	\end{split}
	\ee
	and
	\be
	\eta_N = \frac{g \left(\frac{5}{1-2 \lambda } -\frac{9}{(1-2 \lambda )^2} -7\right)}{3 \pi  \left(1 + \frac{g}{12 \pi } \left(\frac{5}{1-2 \lambda }-\frac{6}{(1-2 \lambda )^2}\right)\right)} \, . 
	\ee
	Here we have specified $d=4$ for explicitness.
	
	We then determine the fixed points of this system. Since the truncation retains a finite number of interaction monomials $\cO_i$ only, this search turns into the algebraic problem of finding the roots of the system $\{\beta_\lambda = 0, \beta_g = 0\}$, cf.\ Table \ref{tab.diffeqs}. Inspecting \eqref{beta-litim}, one finds two fixed points
	\be\label{fp.pos}
	\begin{array}{ll}
		\text{GFP:} & \quad \{g_* = 0, \; \lambda_* = 0\} \, ,  \\[1.3ex]
		\text{NGFP:} & \quad  \{g_* = 0.707, \; \lambda_* = 0.193 \} \, . 
    \end{array}
    \ee
	These correspond to a free and interacting theory, respectively. The NGFP is \emph{the projection} of the Reuter fixed point onto the subspace spanned by the ansatz \eqref{eq.einsteinhilbert}. 
	
	The stability properties of the RG flow in the vicinity of these fixed point are readily obtained by evaluating the stability matrix \eqref{def.B} for the beta functions \eqref{beta-litim}. This yields
	\be\label{fp.stab}
	\begin{array}{ll}
		\text{GFP:} & \quad \{\theta_1 = 2, \; \theta_2 = -2\} \, ,  \\[1.3ex]
		\text{NGFP:} & \quad  \{\theta_{1,2} = 1.48 \pm 3.04 i \} \, . 
	\end{array}
	\ee
	Thus the GFP constitutes a saddle point with one UV-attractive and one UV-repulsive eigendirection. Analyzing the corresponding eigenvectors shows that RG trajectories with a non-vanishing Newton's coupling are repelled by the GFP as $k\rightarrow\infty$. Hence this fixed point cannot act as the UV-completion of gravity. In contrast, the NGFP is UV-attractive for both $g_k$ and $\lambda_k$. The complex stability coefficients indicate that the RG flow spirals into the fixed point as $k \rightarrow \infty$. Thus this fixed point acts as UV-completion for the RG trajectories entering its vicinity.

	\begin{table}[t!]\centering
		\begin{minipage}[c]{0.48\linewidth}
			\centering
			\renewcommand{\arraystretch}{1.5}
			\begin{tabular}{cccccc}
				\hspace{8mm}		& \hspace{2mm}	$d$ \hspace{2mm} & \hspace{2mm} $g_*$ \hspace{2mm} & \hspace{2mm} $\lambda_*$ \hspace{2mm} & \hspace{2mm} $\theta_1$ \hspace{2mm} & \hspace{2mm} $\theta_2$ \hspace{2mm} \\ \hline \hline
				GFP & $d$ & $0$ & $0$ & $2-d$ & $2$ \\		
				\hline
				NGFP &		$2+\epsilon$   &   $\frac{3}{38} \epsilon$   &  $-\frac{3}{38} \epsilon$   &   $\epsilon$   &   $2 + \frac{1}{19} \epsilon$  \\
				NGFP &		$3$ & $0.20$ & $0.06$ & \multicolumn{2}{c}{$1.15 \pm 0.83i$} \\
				NGFP &		$4$ & $0.71$ & $0.19$ & \multicolumn{2}{c}{$1.48 \pm 3.04i$} \\
				NGFP &		$5$ & $2.85$ & $0.24$ & \multicolumn{2}{c}{$2.69 \pm 5.15i$} \\ \hline \hline \vspace{2mm}
			\end{tabular} 
		\end{minipage}
		\begin{minipage}[c]{0.48\linewidth}
			\centering
			\includegraphics[width=52mm]{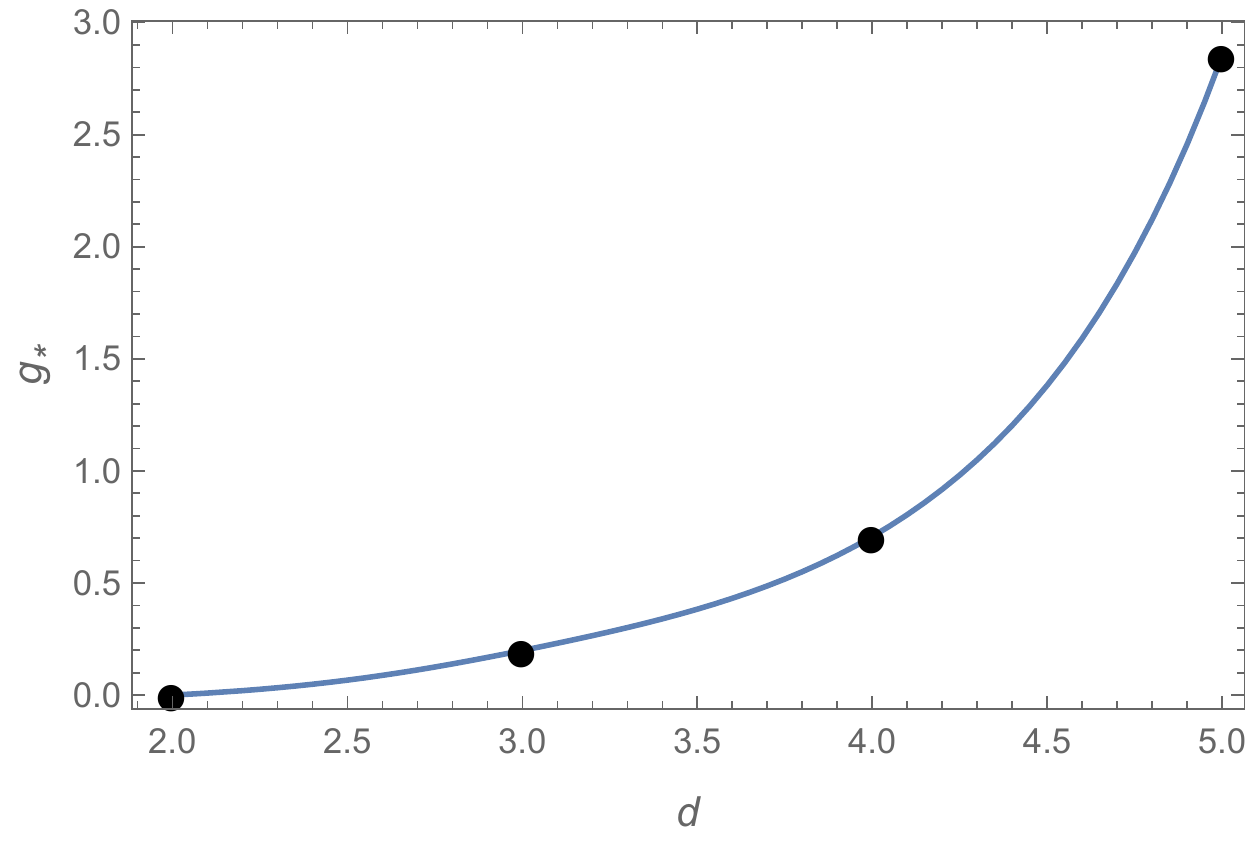}
		\end{minipage}
	\caption{\label{tab.NGFPd} Characteristics of the family of NGFPs in various dimensions $d$ \cite{Souma:1999at,Reuter:2001ag}. The table on the left gives the position and stability coefficients of the fixed points for selected values of $d$. The diagram to the right displays that the family emerges from the Gaussian fixed point in $d=2+\epsilon$ and continuously connects to the Reuter fixed point in $d=4$. Whether there is an upper critical dimension where the family of fixed points ceases to exist is currently an open question. }
\end{table}
	Treating the dimension $d$ as a continuous parameter, one can trace the properties of the NGFP when performing an analytic continuation of the spacetime dimension. The results are summarized in Table \ref{tab.NGFPd}. The table in the left panel gives the position and stability coefficients of the NGFP for selected values $d$ while the diagram in the right panel shows the position $g_*(d)$. The latter illustrates that the family of NGFPs emerges from the GFP in $d=2+\epsilon$ dimensions. It can then be analytically continued up to $d=4$. Thus the Reuter fixed point is the analytic continuation of the NGFP seen in the $\epsilon$-expansion around the free theory at the lower critical dimension $d=2$ \cite{Souma:1999at,Reuter:2001ag}. For $d > 4$, the system \eqref{beta-litim} suggests that the NGFP continues to exist for all dimensions $d > 2$ \cite{Litim:2003vp}. At $d \gtrsim 5$, the existence of the fixed point turns into a regulator-dependent statement though \cite{Reuter:2001ag}. Hence, it is currently unclear if there is an upper critical dimensions where the family of NGFPs ceases to exist.
	
\begin{figure}[t!]
	\centering
	\includegraphics[width=.9\textwidth]{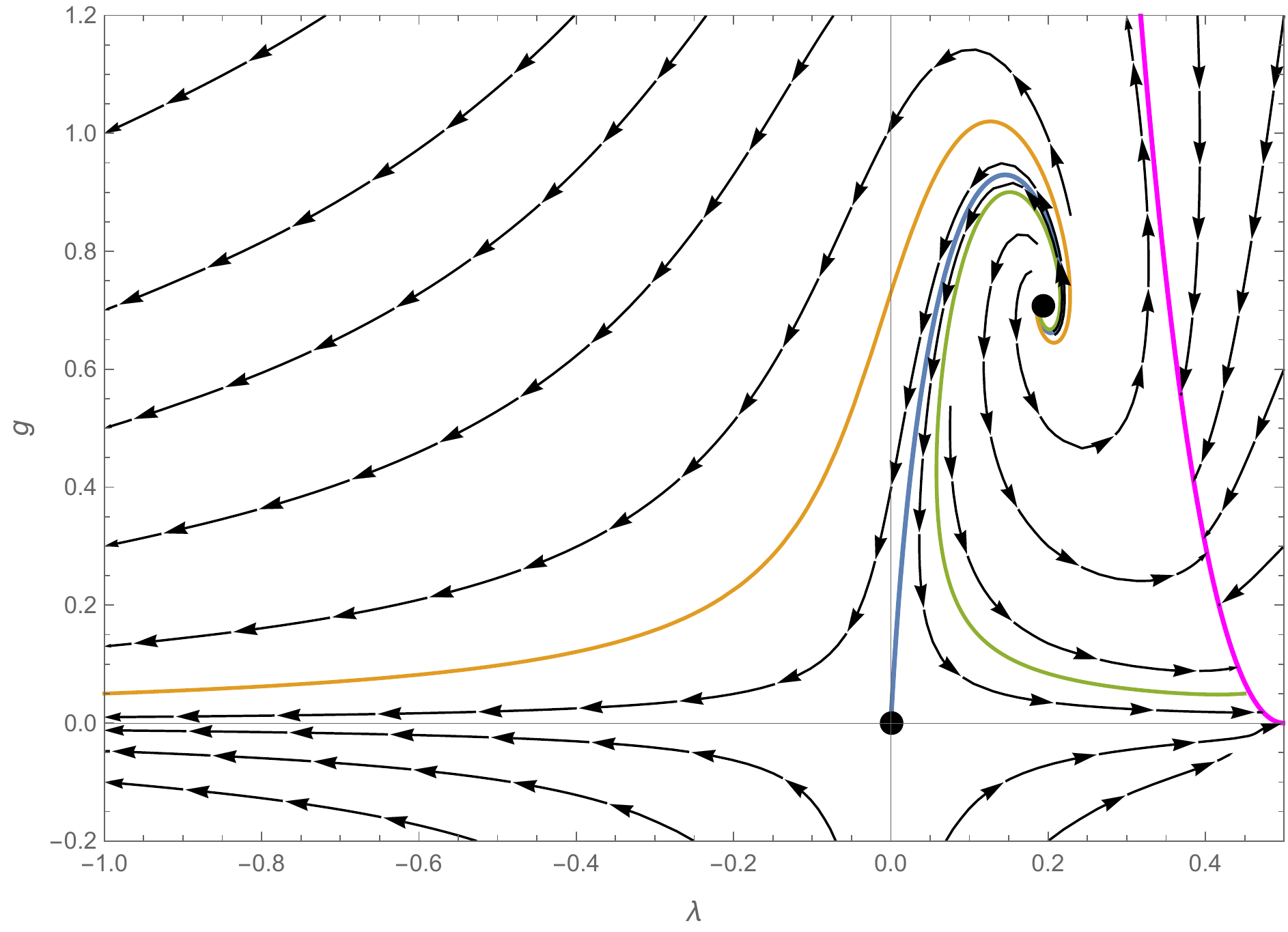}
	\caption{\label{fig:phasespace} Phase diagram constructed from integrating the beta functions \eqref{eq.betadef} in $d=4$. All arrows point towards lower coarse-graining scale $k$. The GFP and NGFP are marked by the black dots while the magenta line displays the locus where the anomalous dimension $\eta_N$ diverges. The NGFP acts as an UV-attractor capturing all trajectories in its vicinity. Lowering the coarse-graining scale, the flow undergoes a crossover towards the GFP. The separatrix connecting the two fixed points is highlighted in blue. This solution leads to a vanishing cosmological constant $\Lambda_0 = 0$. Trajectories to the left of this line, exemplified by the orange trajectory, have been classified as Type Ia  and are characterized by $\Lambda_0 < 0$. RG trajectories to its right constitute the Type IIIa solutions (represented by the green trajectory). They terminate at a finite value of $k$ and lead to positive values $\Lambda_k > 0$.  (Initially constructed in \cite{Reuter:2001ag}).}
\end{figure}

	The system \eqref{beta-litim} is readily integrated numerically. The resulting phase diagram is governed by the interplay of the fixed points \eqref{fp.pos} and shown in Fig.\ \ref{fig:phasespace}.
Here the magenta line indicates the position of a singular locus where $\eta_N$ diverges. The physically relevant part of the phase diagram consists of the RG trajectories with emanate from the NGFP in the UV and cross over to the GFP as $k$ decreases. A special role is thereby played by the separatrix (blue line) connecting the two fixed points. This trajectory leads to a vanishing cosmological constant $\lim_{k\rightarrow 0} \Lambda_k = 0$. The trajectories to the right of this line are classified as Type Ia (orange line). Their characteristic feature is a negative cosmological constant, $\lim_{k\rightarrow 0} \Lambda_k < 0$. The trajectories to the right of the separatrix (green line) have been labeled Type IIIa. They terminate at the singular locus at a finite value of $k$. In the vicinity of the GFP they exhibit a regime where $\Lambda_k$ is constant and positive. It is expected that nature is described by an RG trajectory within this class \cite{Reuter:2004nx}. This trajectory is special in the sense that it almost hits the GFP. Only at the very last moment, it takes a turn flowing away from the fixed point. In this way the trajectory accommodates the tiny value of the cosmological constant observed in cosmology. Thus, the Einstein-Hilbert truncation does not predict the value of the cosmological constant. It is a function of the free parameters labeling the RG trajectories leaving the NGFP. The cosmological constant then has the role of an experimental input which is used to identify the RG trajectory realized in nature.

\begin{figure}[t!]
	\centering
	\includegraphics[width=.45\textwidth]{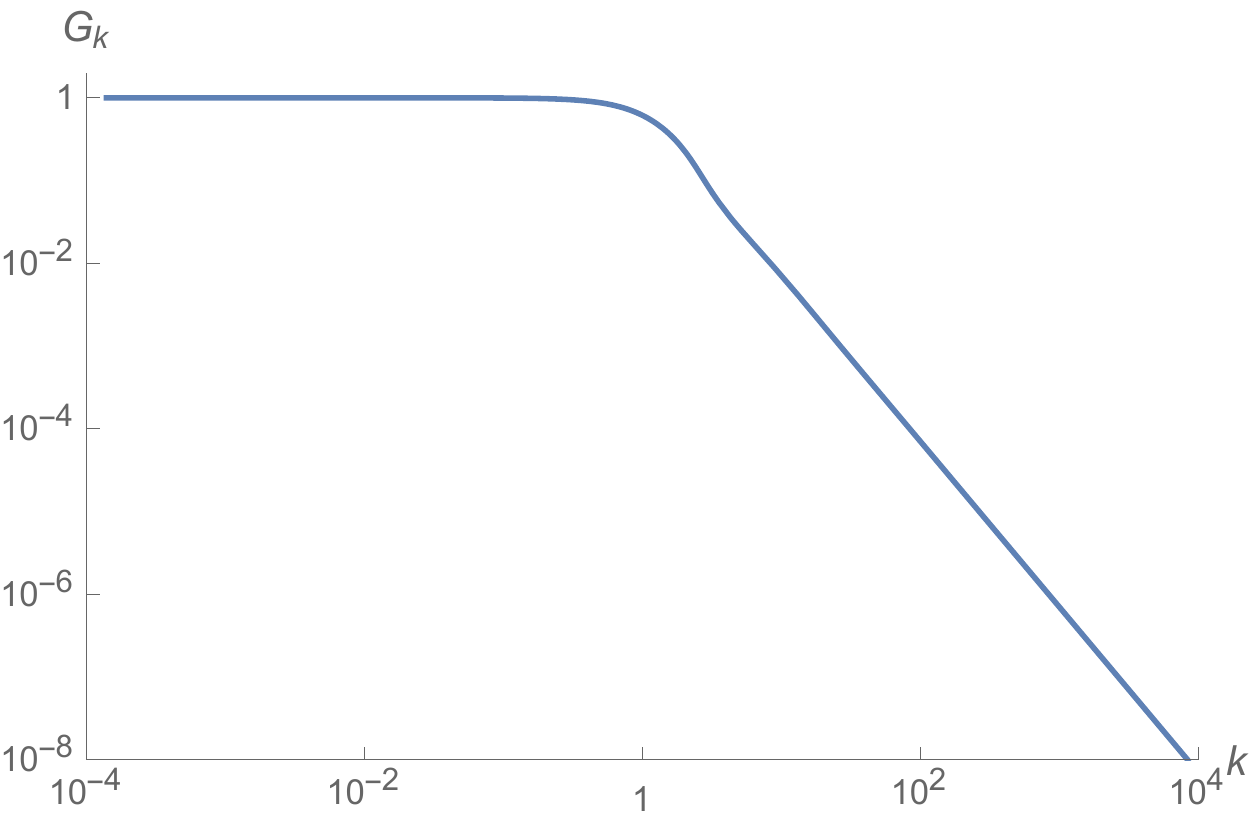}
	\includegraphics[width=.45\textwidth]{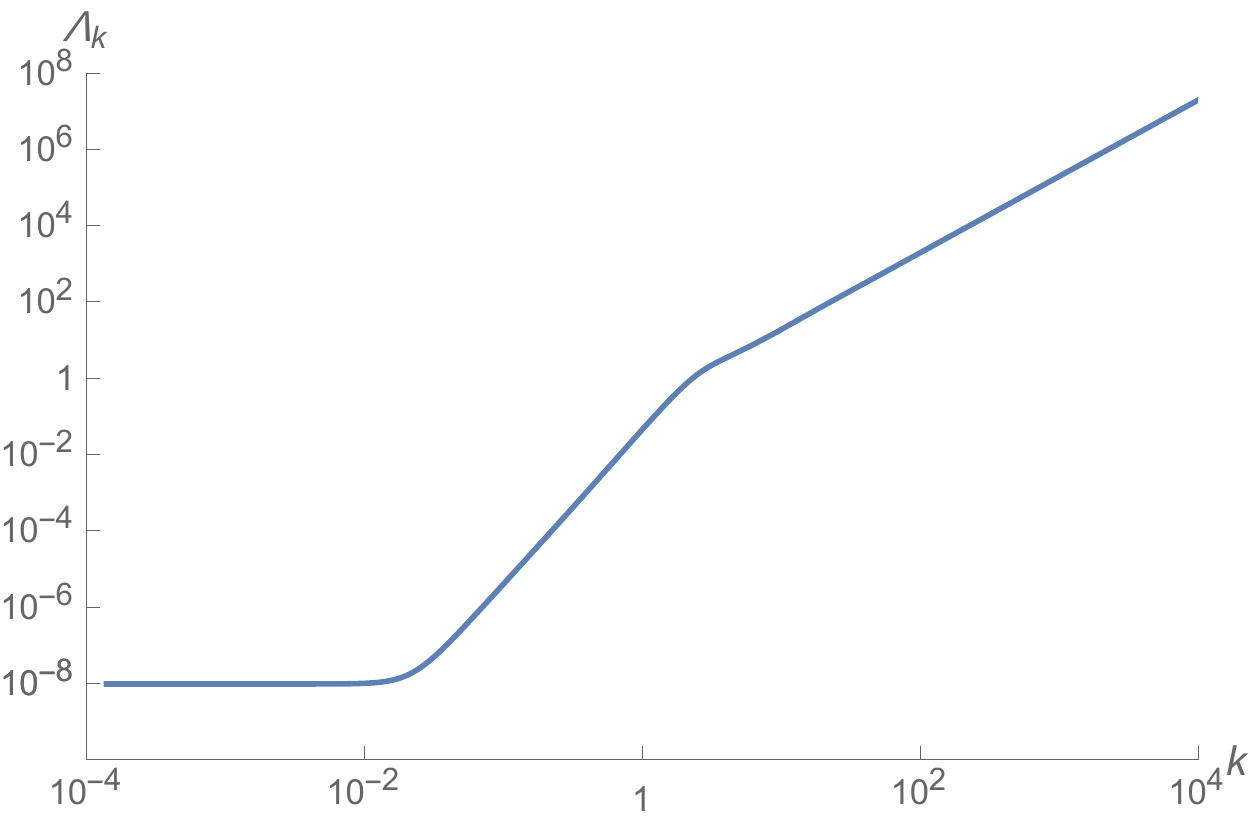}
	\caption{\label{fig:coarsegrainingdimful}  Dependence of the dimensionful Newton's coupling (left panel) and cosmological constant (right panel) on the coarse-graining scale along a typical RG trajectory of Type IIIa. The flow interpolates between the classical regime ($k \ll 1$) where $G_k$ and $\Lambda_k$ are constant and the fixed point regime ($k \gg 1$) where $G_k \propto k^{-2}$ and $\Lambda_k \propto k^2$. By definition, the cross-over between these scaling regimes occurs at the Planck scale $M_{\rm Pl} \equiv G_0^{-1/2}$ which is generated dynamically when flowing away from the NGFP. Notably, $\Lambda_k$ exhibits an intermediate scaling regime where $\Lambda_k \propto k^4$. All quantities are measured in units of $M_{\rm Pl}$. (Adaptation from \cite{Gubitosi:2018gsl}).}
\end{figure}
At this stage, it is instructive to pick a generic RG trajectory of Type IIIa and illustrate the $k$-dependence of the dimensionful couplings. The resulting flow of $G_k$ and $\Lambda_k$ is exemplified in Fig.\ \ref{fig:coarsegrainingdimful} where all dimensionful quantities are given in units of the Planck scale $M_{\rm Pl} = G_0^{-1/2}$. For $k > 1$ the scale-dependence is governed by the NGFP while for $k < 1$ the flow is controlled by the GFP. As a result, the flow of the couplings interpolates between the scaling regimes
\be\label{fp.crossover}
\begin{array}{llll}
	\text{NGFP:} & \quad  G_k \simeq g_* k^{-2} \, , & \quad \Lambda_k \simeq \lambda_* k^2 & \quad k > 1 \, , \\[1.2ex]
	\text{GFP:} & \quad G_k \simeq G_0 \, , & \quad \Lambda_k \simeq \Lambda_0 & \quad k < 1 \, . 
\end{array}
\ee
The crossover between the two regimes occurs at the Planck scale. This scale is generated dynamically when moving away from the NGFP. Classical general relativity (in the sense of a low-energy effective field theory) is then recovered in the vicinity of the GFP.

We conclude by stressing that the exact fixed point action $\Gamma_*$ associated with the Reuter fixed point does (most likely) not coincide with the Einstein-Hilbert action. While this may be suggested by the analysis of this section, one has to account for the fact that we have been working within a projection of the full theory space to this two-dimensional subspace. Additional contributions to $\Gamma_*$, as, e.g., higher-derivative terms, are not visible in this analysis.  

\subsection{Further reading}
\label{sec.further-reading}
The Einstein-Hilbert truncation discussed in this section constitutes the starting point for understanding the theory space of gravity, its RG fixed points, and their mutual relations. By now, this exploration has made significant progress in moving beyond this basic example. At the level of the background approximation $f(R)$-type projections have been studied in polynomial approximations to very high order \cite{Codello:2007bd,Machado:2007ea,Falls:2013bv,Falls:2017lst,Falls:2018ylp} and it has been shown that the NGFP also persists once the two-loop counterterm identified by Goroff and Sagnotti \cite{Goroff:1985sz,Goroff:1985th} is included in the projection \cite{Gies:2016con}. Within the fluctuation approach there has been significant progress on understanding the momentum-structure of the graviton propagator \cite{Christiansen:2014raa,Bonanno:2021squ} and resolving the momentum-dependence of three- and four-point vertices \cite{Christiansen:2015rva,Denz:2016qks}. In parallel, a program geared towards developing asymptotically safe amplitudes has been initiated in \cite{Draper:2020bop}. Covering these developments in detail is beyond the scope of this introductory chapter and the interested reader may consult the more advanced chapters of this volume for further information.
	\section{Concluding comments}
	\label{sec.conclusion}
The Wetterich equation \eqref{eq.Wetterich} constitutes an essential tool in developing the gravitational asymptotic safety program. Starting from its adaption to gravity \cite{Reuter:1996cp}, it has provided substantial evidence for the existence of a viable interacting renormalization group fixed point -- the Reuter fixed point -- which could provide a consistent and predictive high-energy completion of the gravitational interactions. 

The present chapter focused on the case where the gravitational degrees of freedom are carried by the metric field. The applicability of the functional renormalization group and in particular the Wetterich equation is not limited to this setting though. It has readily been extended to other sets of fields which, at the classical level, encode the same gravitational dynamics as general relativity. Notably, this includes the case where the gravitational degrees of freedom are encoded in the vielbein (``tetraed only''-formulation) \cite{Harst:2012ni,Dona:2012am}, the Palatini formalism \cite{Harst:2014vca,Harst:2015eha,Pagani:2015ema,Gies:2022ikv}, the Arnowitt-Deser-Misner decomposition \cite{Manrique:2011jc,Rechenberger:2012dt,Biemans:2016rvp,Biemans:2017zca,Houthoff:2017oam}, and unimodular gravity \cite{Eichhorn:2013xr,Eichhorn:2015bna,Percacci:2017fsy,deBrito:2020xhy}. While the exploration of the corresponding theory spaces is far less developed than the one for the metric theory, there are indications that these settings also possess interacting renormalization group fixed points suitable for rendering the construction asymptotically safe. In the case of unimodular gravity, there are arguments that the theory is in the same universality class as the metric formulation \cite{deBrito:2021pmw}. Whether the other fixed points are in the universality class of the Reuter fixed point is an open question though. 

Notably, there also has been progress aiming at the implementation of the renormalization group on discrete geometries. In the context of the Causal Dynamical Triangulation program \cite{Ambjorn:2012jv,Loll:2019rdj}, renormalization group flows have been constructed in \cite{Ambjorn:2014gsa}. The underlying idea is to pick an observable whose value is held constant when varying the parameters of the Monte Carlo simulation. This led to the surprising conclusion that the phase-transition line expected to provide the high-energy completion of the theory actually appears to be approached in the infrared. 

So far, our discussion has focused on gravitational degrees of freedom only. It is rather straightforward to extend this construction by including additional matter fields as well as all the building blocks of the standard model of particle physics \cite{Eichhorn:2018yfc,Eichhorn:2022gku}. Many of the gravity-matter systems investigated to date exhibit interacting renormalization group fixed points whose properties are very similar to the ones found for the Reuter fixed point. Since the beta functions encoding the fixed point structure of these systems depend on the number of matter fields in a continuous way, it is likely that the Reuter fixed point is part of a continuous web of interacting fixed points. Since it is unlikely that these encode the same universality class, it is suggestive to refer to these as deformed Reuter fixed points, highlighting that the systems actually realize different (albeit related) universal behaviors. A detailed summary of the state-of-the-art in investigating asymptotically safe gravity-matter systems is beyond the scope of this elementary introduction and we refer to the recent reviews \cite{Eichhorn:2018yfc,Eichhorn:2022gku} as well as other chapters of this book. In short, it is conceivable though that the asymptotic safety mechanism may lead to a unified theory incorporating the standard model of particle physics and gravity within the framework of a relativistic quantum field theory. This exciting perspective certainly warrants further investigation. 

	\section*{Acknowledgements}
My understanding of the functional renormalization group and its applications in the context of gravity has benefited enormously from countless discussions with many colleagues. It is therefore my pleasure to thank 
J.\ Ambj{\o}rn, 
D.\ Becker, 
W.\ Beenakker,
A.\ Bonanno,
L.\ Bosma,
T.\ Budd,
L.\ Buoninfante, 
J.\ Donoghue, 
T.\ Draper,
A.\ Eichhorn,
R.\ Ferrero, 
G.\ Gubitosi,
R.\ Kleiss,
A.\ Koshelev, 
R.\ Loll,
M.\ Niedermaier,
R.\ Ooijer,
C.\ Pagani, 
J.\ M.\ Pawlowski, 
R.\ Percacci, 
A.\ D.\ Pereira, 
S.\ Pirlo,
A.\ Platania, 
M.\ Reichert,  
M.\ Schiffer, 
O.\ Zanusso,
and C.\ Wetterich for sharing their insights and views. In addition, I want to thank M.\ Becker and A.\ Ferreiro for their insightful comments on the manuscript and B.\ Knorr and C.\ Ripken for their close collaboration on many aspects presented in this review. Finally, I would like to thank M.\ Reuter for introducing me to this subject and his continual advice and support.


\providecommand{\href}[2]{#2}\begingroup\raggedright\endgroup
\end{document}